\documentclass[10pt,aps,twocolumn,nofootinbib,superscriptaddress,longbibliography]{revtex4-2}

\usepackage{graphicx}% Include figure files
\usepackage{amsmath}
\usepackage{dcolumn}% Align table columns on decimal point
\usepackage{bm}% bold math/
\usepackage{hyperref}% add hypertext capabilities
\newcommand{\ee}{\text{e}}
\newcommand{\dd}{\text{d}}
\newcommand{\prefactor}{\psi}
\usepackage[dvipsnames]{xcolor}

\begin{document}

\title{Hidden universality of rare fluctuations and a central-limit structure at criticality}

\author{N.~G.~Fytas}
\affiliation{School of Mathematics, Statistics and Actuarial Science, University of Essex, Colchester CO4 3SQ, United Kingdom}
\affiliation{Section of Condensed Matter Physics, Department of Physics, National and Kapodistrian University of Athens, Panepistimiopolis, 157 72 Athens, Greece}

\author{V.~Martín-Mayor}
\affiliation{Departamento de Física Teórica, Universidad Complutense, 28040 Madrid, Spain}

\author{A.~L.~Stella}
\affiliation{Department of Physics and Astronomy, University of Padova, Via Marzolo 8, I-35131 Padova, Italy and INFN, Sezione di Padova, Via Marzolo 8, I-35131 Padova, Italy}

\author{G.~Teza}
\affiliation{Max Planck Institute for the Physics of Complex Systems, Nöthnitzer Str. 38, 01187 Dresden, Germany}
\affiliation{Dipartimento di Fisica, Università di Trieste, Strada Costiera 11, I-34151 Trieste, Italy}

\author{A.~Vasilopoulos}
\affiliation{School of Mathematics, Statistics and Actuarial Science, University of Essex, Colchester CO4 3SQ, United Kingdom}

\author{D.~Yllanes}
\affiliation{Fundación ARAID, Diputación General de Aragón, 50018 Zaragoza, Spain}
\affiliation{Instituto de Biocomputación y Física de Sistemas Complejos (BIFI), Universidad de Zaragoza, 50018 Zaragoza, Spain}
\affiliation{Zaragoza Scientific Center for Advanced Modeling (ZCAM), 50018 Zaragoza, Spain}

\date{\today}

\begin{abstract}
The central limit theorem is among the most fundamental principles in the natural sciences, governing the fluctuations of sums of weakly correlated random variables through Gaussian statistics. At critical points, however, long-range correlations invalidate this paradigm, raising a fundamental question: does any general limit theorem survive at criticality? Although scale invariance has long been recognised as the defining feature of typical critical fluctuations, the behaviour of rare fluctuations has remained largely unexplored, as their extraordinarily small probabilities place them beyond the reach of conventional numerical methods.
Recent theoretical advances suggest that universal asymptotic laws {may govern large deviations in systems at criticality}, yet direct numerical evidence has been lacking.
Here we provide the first such evidence in the two-dimensional Ising universality class.
Using large-scale Tethered Monte Carlo simulations, we probe order-parameter fluctuations with probabilities as small as $10^{-200}$ and show that, once normalised by their characteristic width, the far tails collapse onto a single asymptotic law shared by distinct microscopic models in the same universality class.
This collapse is governed by a universal amplitude, whose numerical extraction is compatible
with the conjectured exact result we derive from conformal field theory, and points towards a generalised central limit theorem for critical fluctuations.
\end{abstract}
\maketitle

\section{Introduction\label{sec:Intro}}

The central limit theorem for sums of many independent or weakly correlated random variables is one of the most fundamental results in probability theory, governing fluctuations in an extraordinary range of natural phenomena ~\cite{alma991005852086808044}. Central to its broad applicability is a profound universality: once the sums are normalised to account for the number of variables, their finite individual variances, and their correlations, fluctuations about the mean converge to the same Gaussian distribution, irrespective of the underlying microscopic details.

Universality is likewise the central organizing principle of the modern theory of critical phenomena based on the renormalisation group~\cite{kadanoff2000statistical}. It is therefore unsurprising that the central limit theorem served as a natural prototype in early attempts to place the renormalisation-group framework on rigorous probabilistic foundations \cite{cassandro1978critical,jona2001renormalization} . At criticality, however, the strong correlations between microscopic variables invalidate the assumptions underlying the classical theorem and no result of comparable generality has emerged. Exact probability distributions for sums have been obtained only in special settings, such as mean-field~\cite{teza2025universal} and hierarchical models~\cite{jona2001renormalization}. These cases provide valuable insight but leave open the extent to which the resulting distributions are truly universal.

Especially since the pioneering work of Binder in 1981~\cite{binder1981}, probability distributions of sums of many critical variables, such as the magnetisation of a ferromagnet at its critical point, have been the subject of extensive analytical and numerical investigation ~\cite{gallavotti1975,Bruce1995,Hilfer1995,smith1995study,tsypin2000,malakis2006universal,hilfer2003multicanonical,hilfer2005multicanonical,Delamotte,balog2025universal}. Progress toward a complete characterisation of these distributions has proved considerably more challenging than for other universal quantities, such as critical exponents and amplitude ratios~\cite{amit:05}. A key obstacle is that the long-range correlations at criticality make the full probability distributions sensitive to the geometry, topology and boundary conditions of the finite systems from which the thermodynamic limit is approached~\cite{cardy:84b,burkhardt:85,kaneda2001finite}. This raises the possibility that universality may emerge only in a restricted asymptotic regime of the distributions, rather than in their full functional form. Indeed, most studies featuring scale invariance (see, \emph{e.g.},~\cite{amit:05}), and even universality aspects~\cite{bruce1992,wilding:1997}, have been restricted to typical fluctuations. A central, yet unfulfilled aim is to determine whether these distributions are universal across their whole range ---including, crucially, large and rare fluctuations.

The emergence of large-deviation theory~\cite{Touchette2009} has done little to resolve this issue. Although it provides the natural framework for characterising rare fluctuations and their universal properties, it has so far found limited application in critical systems~\cite{stella2023anomalous,teza2025universal}. As a result, a fundamental gap remains in the theory of critical phenomena: despite its many successes, it is still unknown to what degree universality extends to the probability distributions of critical fluctuations, or whether they obey a principle analogous to the central limit theorem.

Early studies~\cite{Bruce1995,Hilfer1995} identified the far tails of critical probability distributions as the natural place to search for universal behaviour, motivated by the conjectured ---and in some cases rigorously established~\cite{mccoy1973two,camia2016planar,camia2025power}---  connection between their asymptotic form and the critical equation-of-state exponent. Yet these tails proved extraordinarily difficult to investigate numerically~\cite{Hilfer1995,smith1995study,tsypin2000,malakis2006universal}. Critical slowing down severely reduces the statistical efficiency of large-scale simulations, while the overwhelming rarity of the relevant fluctuations places the asymptotic regime far beyond the reach of conventional Monte Carlo methods. Indeed, it was argued that these tails were effectively inaccessible with the most advanced simulation techniques then available~\cite{hilfer2003multicanonical,hilfer2005multicanonical}.

\begin{figure*}
\centering
\includegraphics[width=\linewidth]{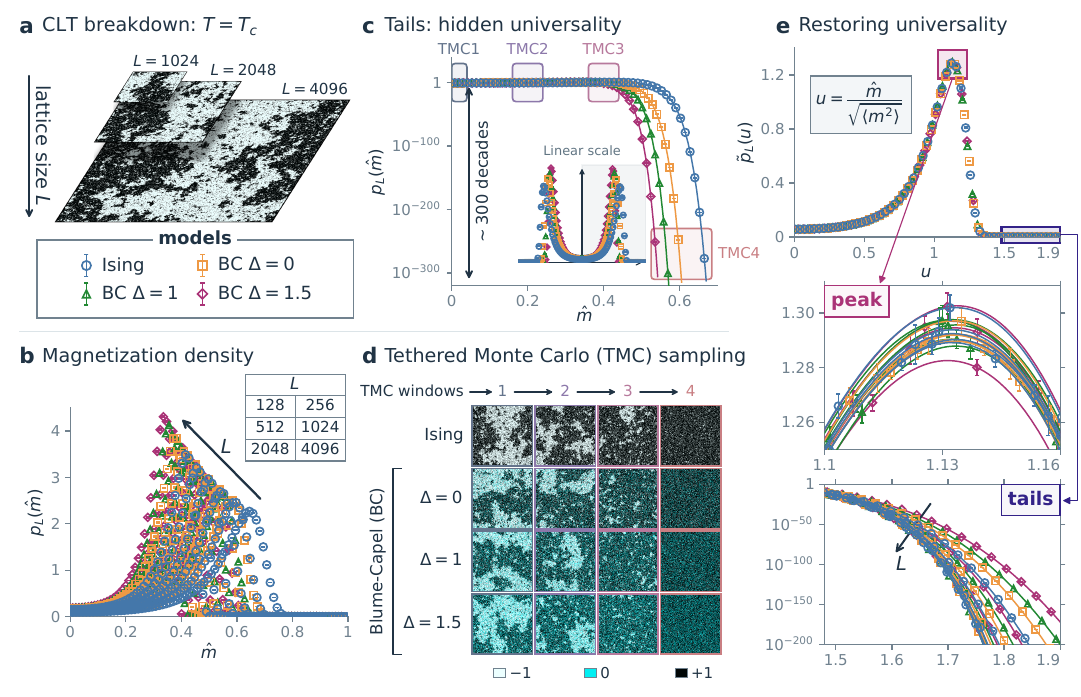}
\caption{\textbf{Hidden universality of rare fluctuations at 
criticality.} Panel~{\bf a} depicts the breakdown of the central limit theorem: at criticality, $T=T_\mathrm{c}$, spin configurations exhibit scale invariance at all length scales as the system size is increased, a signature of the long-range correlations that invalidate the theorem's assumptions. Although in panel~{\bf a} we show configurations for the Ising model, we consider four different 
models belonging to the same universality class: the Ising model on the square lattice and three instances of the Blume--Capel (BC) model~\cite{blume,capel6} with crystal-field $\varDelta=0$, $1$ and $1.5$ (see Sec.~\ref{subsect:models} in {\bf Methods} for definitions). The probability density of the order parameter varies significantly with system size $L$, and differs markedly across our four models (panel~{\bf b}; system sizes ranging from $L=128$ to $L=4096$). Although scale invariance for typical fluctuations is restored with the rescaling indicated in the box, we are primarily interested in the tails of the distribution (panel~{\bf c}). The Tethered Monte Carlo (TMC) method~\cite{fernandez:09,martin-mayor:11} gives us access to these tails: universality is reflected in the structural similarity of spin configurations spanning the same probability ranges across our four models (panel~{\bf d}; probability ranges are indicated by the boxes in panel~{\bf c}). The universality of the order-parameter distribution is illustrated in panel~{\bf e}, where the order parameter is expressed in units of the distribution width. For typical fluctuations (probabilities of order one), universality is apparent already for very small $L$. However, recovering universality for probabilities as small as $10^{-200}$ requires simulating sufficiently large systems. Error bars are one standard deviation.
\label{fig:graphical-abstract}}
\end{figure*}

However, although from the numerical side the situation has been stalled for decades, a recent analytical advance by Teza and Stella~\cite{teza2025universal} has reframed the problem entirely. They showed that the extensivity of thermodynamic potentials imposes remarkably strong constraints on the asymptotic form of critical probability distributions. In particular, the leading asymptotic structure is universal except for a single amplitude associated with the normalisation of the fluctuating variable.
This theoretical picture has recently received numerical support through studies of the asymptotic tails in several critical models~\cite{zhu2026}. 
Yet whether this remaining amplitude becomes universal after 
appropriate normalisation has remained an open question.
The amplitude is directly related to {that} governing the singular part of the thermodynamic free-energy density, and therefore represents the only possible source of non-universality in the far tails.
Since it rescales trivially under a change of normalisation, testing universality reduces to two tasks: identifying the appropriate 
normalisation and determining the tails with sufficient numerical precision. Strong evidence points to the width of the magnetisation distribution as the natural normalisation~\cite{binder1981,smith1995study,Hilfer1995,%
hilfer2003multicanonical,bruce1992}, a conclusion supported by 
Salas and Sokal~\cite{salas:00}, whose comparison of conformal 
field theory (CFT) predictions with extensive Monte Carlo simulations 
confirms universality ---but only for fluctuations with 
probabilities no smaller than $\sim 10^{-6}$.

Here, we confirm universality over a vastly extended range of 
probabilities, as small as $10^{-200}$, by means of the Tethered Monte Carlo method (TMC, see Sec.~\ref{subsect:TMC-main}), a specialised simulation technique that gives direct access to the far tails of critical probability distributions. To demonstrate universality, we apply this approach to four distinct realisations of the square-lattice Blume--Capel (BC) model~\cite{blume,capel6}, including the Ising model as a limiting case; see Sec.~\ref{subsect:models}. These systems belong to the same universality class, sharing the same critical exponents and universal amplitude ratios while differing in their microscopic interactions, and therefore provide a stringent test of whether the asymptotic structure of critical fluctuations is genuinely 
universal. The detailed theoretical predictions against which our data are compared are summarised in 
Sec.~\ref{subsect:theoretical-predictions}. Strikingly, in contrast with typical fluctuations, universality in the asymptotic regime becomes evident only for sufficiently large system sizes, as illustrated in Fig.~\ref{fig:graphical-abstract}. 
{Our study of the asymptotic regime will not focus only on the leading exponential decay, but it will also include the algebraic prefactor.}

\subsection{Tethered Monte Carlo: Overview \label{subsect:TMC-main}}

Accessing the regime of large fluctuations is a major computational challenge.
Observing an event of probability $p$  even once requires, on average, $p^{-1}$ statistically 
independent samples, placing probabilities of order $10^{-100}$ and below far beyond the reach of brute-force sampling. 
Conventional Monte Carlo simulations typically yield only about 
$10^6$ independent samples, making direct estimation of such 
rare events infeasible. Indeed, the probabilities explored in 
recent work are not smaller than $10^{-5}$~\cite{zhu2026}. 
Overcoming this limitation requires a fundamentally different 
strategy: a Monte Carlo method that samples the logarithm of 
the probability rather than the probability itself. 
TMC~\cite{fernandez:09}, an advanced form of the microcanonical 
Monte Carlo method of Ref.~\cite{martin-mayor:07}, is ideally 
suited to this task. Combined with a cluster update 
algorithm~\cite{martin-mayor:09}, which substantially mitigates 
critical slowing down, it enables high-precision simulations of 
square lattices with linear sizes up to $L=4096$ at the critical temperature, allowing us to resolve order-parameter probabilities down to $10^{-200}$ (Fig.~\ref{fig:graphical-abstract}-{\bf c}).

To define the TMC formalism ---see \textbf{Methods} and 
Refs.~\cite{martin-mayor:11,yllanes:11} for details--- we start 
from a micromagnetic ensemble and couple the system to a 
magnetostat (an external device that fixes the mean 
magnetisation, in analogy with a thermostat). {We thus obtain a 
tethered ensemble in which the pdf $p_L(\hat{m})$ of the tethered 
magnetisation $\hat{m}$ arises as a Gaussian convolution of the 
canonical probability density $\rho_L(m)$ of the magnetisation per spin $m$; see 
Sec.~\ref{subsect:convolution} in \textbf{Methods} for details. 
Monte Carlo simulations are then carried out at fixed values of 
$\hat{m}$, measuring the tethered field $\hat{h}$,
\begin{equation}\label{eq:hath}
\langle \hat{h}\rangle_{L,\hat{m}} = 
\langle \hat{m} - m\rangle_{L,\hat{m}}= 
- \frac{1}{N}\frac{\dd \log p_L(\hat{m})}{\dd \hat{m}}\,.
\end{equation}
In the $L\to\infty$ limit this fluctuation-dissipation relation gives direct access to 
the full probability density $p(\hat{m})$, including its
far tails.}

\subsection{Universal amplitude and asymptotic law\label{subsect:theoretical-predictions}}

Here we summarise the theoretical predictions to be tested in our simulations.
{Throughout, each model is considered at its critical temperature, which is omitted from expectation-value subscripts.
}

{We compare different models belonging to the same universality class, all defined on square lattices of linear size $L$ with periodic boundary conditions (Fig.~\ref{fig:graphical-abstract}-{\bf b}).
Maintaining the geometry and boundary conditions is crucial~\cite{cardy:84b,burkhardt:85}.
To eliminate non-universal magnetisation scales, we use $u=\hat{m}/\sqrt{\langle m^2\rangle_L}$, where the denominator is the zero-field canonical rms magnetisation.}

In the $L\to\infty$ limit, at the critical temperature, the probability density function $\tilde{p}(u)$ at large $|u|$ behaves as $\log \tilde{p}(u) \sim -c^* |u|^{\delta+1}$, irrespective of the model under consideration. Here, $\delta=y_h/(d-y_h)$ is the universal equation-of-state exponent ($d$ is the space dimension; for the $d=2$ Ising universality class, $y_h=15/8$ and $\delta=15$). Since we directly measure the derivative of 
the logarithm of the probability, Eq.~\eqref{eq:hath}, the 
following asymptotic expansion is more convenient to test:
\begin{equation}\label{eq:universality}
S(u)=\lim_{L\to\infty}-\frac{\mathrm{d}\log\tilde{p}_L(u)}{\mathrm{d} u}\sim c^{**}  u^\delta\,,\,c^{**}=(\delta+1)c^*\,.
\end{equation}
Our primary goal is to determine the amplitude $c^{**}$, 
verifying that a \emph{single} value accounts for the results 
from different models in the same universality class. Note that 
the universality of $c^*$ follows from that of $c^{**}$ 
and $\delta$.

A second testable prediction emerged when deriving 
Eq.~\eqref{eq:universality}: a universal combination of scaling 
amplitudes involving a finite-size-scaling amplitude and a 
thermodynamic-limit amplitude. Universal amplitude combinations 
in either the finite-size scaling regime or the thermodynamic 
limit are well known, but Eq.~\eqref{eq:amplitudes} below is, 
to our knowledge, the only example of mixed nature.

On the one hand, we have the amplitude $B$ for the fluctuations 
of the magnetisation at zero magnetic field,
\begin{equation}\label{eq:B-def}
\sqrt{\langle m^2\rangle_L}=
B L^{-\beta/\nu} \,[1+\cdots]\,,
\end{equation}
where $\beta$ is the order parameter critical exponent, $\nu$ 
is the thermal critical exponent ---in our case, $\beta=1/8$ 
and $\nu=1$--- and the dots stand for scaling corrections 
subleading in the limit of large $L$. On the other hand, 
amplitude $A$ governs the magnetisation response to a magnetic 
field $h$ in the thermodynamic limit 
($\tilde{h}=h/T_\mathrm{c}$):
\begin{equation}\label{eq:A-def}
\langle m\rangle_{L=\infty} = A\, 
\frac{\delta+1}{\delta}\, \tilde{h}^{1/\delta}\,+\,\cdots
\end{equation}
where the dots stand for corrections subleading when 
$\tilde{h}\to 0^+$. These two amplitudes are related to 
$c^{**}$ through
\begin{equation}\label{eq:amplitudes}
c^{**}=\left[\frac{\delta}{(\delta+1)}  
\frac{B^{1+\frac{1}{\delta}}}{A}\right]^\delta\,.
\end{equation}
Building on results from Ref.~\cite{teza2025universal}, we derive the leading tail law and Eq.~\eqref{eq:amplitudes} in Sec.~\ref{subsec:FSS} of \textbf{Methods}.

We observe in Sect.~\ref{subsec:CFT-amplitude} that a simplifying assumption allows us
to combine  CFT results from Refs.~\cite{salas:00} and~\cite{caselle:00b},  yielding the following  conjectured exact result  for the square geometry with periodic boundary conditions: 
\begin{equation}\label{eq:c-CFT}
c^{**}_{\mathrm{CFT}}\approx 0.855\,598,
\end{equation}
which will provide a benchmark to our numerical calculations of amplitudes $A$ and $B$.

For this periodic geometry, an algebraic prefactor $\tilde{p}(u)\sim C|u|^\prefactor\exp[-c^{**}|u|^{16}/16]$, with $\prefactor=(\delta-1)/2=7$, has also been predicted~\cite{Bruce1995,Hilfer1995,smith1995study,camia2025power}.
For positive $u$, the corresponding logarithmic derivative has the form
\begin{equation}\label{eq:tail-correction}
S(u)=c^{**}u^{15}-\frac{\prefactor}{u}+\cdots,
\end{equation}
inclusive of a subleading correction with prediction $\prefactor=(\delta-1)/2=7$.
We examine this additional prediction for the same fixed geometry and boundary conditions.

\section{Results\label{sec:results}}

\begin{figure}
\centering
\includegraphics[width=\linewidth]{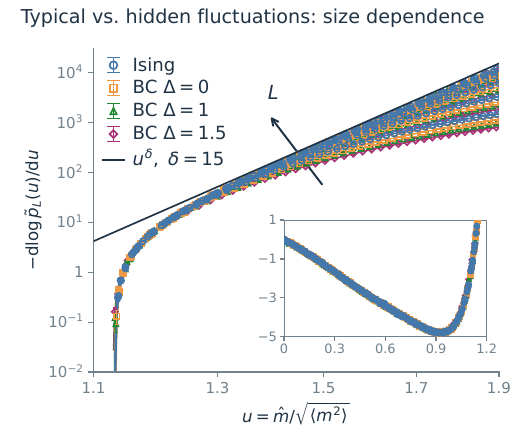}
\caption{\textbf{\boldmath Size dependence of 
$-\mathrm{d}\log\tilde{p}_L/\mathrm{d}u$}. The derivative 
of the logarithm of the probability density 
$-\mathrm{d}\log\tilde{p}_L/\mathrm{d}u$ is shown as a 
function of the dimensionless magnetisation $u=\hat{m}/\sqrt{\langle m^2\rangle_L}$, as computed for our 
four models ---the Ising model and three instances of the 
Blume--Capel (BC) model---, in lattices with linear size $L=128$, $256$, $512$, $1024$, $2048$, and $4096$ (arrows indicate growing 
$L$). For typical fluctuations ({\bf inset}; note the 
different horizontal scale), universality is apparent from 
the collapse onto a master curve of data from all four 
models, for all $L$ values. Instead, for the 
large-deviations region $u > 1.1$, the model-independent 
regime is reached only in the thermodynamic limit 
$L\to\infty$; compare the horizontal scales with 
Fig.~\ref{fig:graphical-abstract}-{\bf e}. The curves 
approach an enveloping curve growing as $u^{\delta}$, 
see Eq.~\eqref{eq:universality}. Error bars are one 
standard deviation.\label{fig:h}}
\end{figure}

\begin{figure}
\centering
\includegraphics[width=\linewidth]{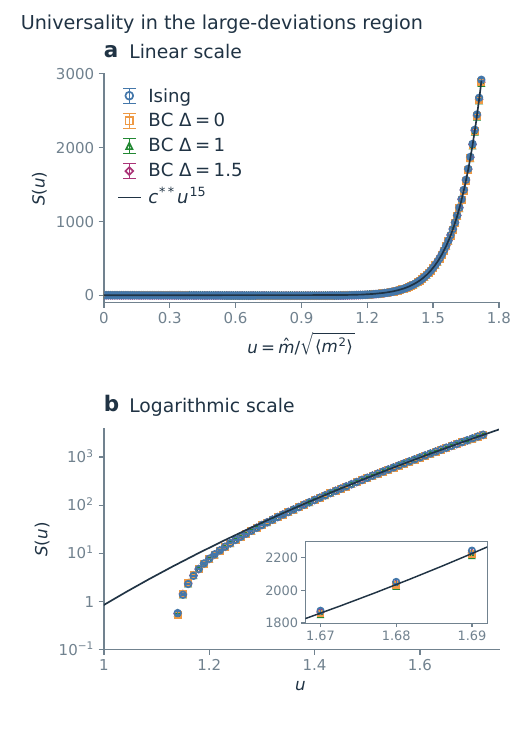}
\caption{\textbf{\boldmath Universality in the large-deviations 
region is restored in the thermodynamic limit.} Infinite-size 
extrapolation $S(u)=\lim_{L\to\infty}[-\mathrm{d}\log
\tilde{p}_L/\mathrm{d}u]$ versus $u$, as obtained from our 
four models using Eq.~\eqref{eq:extrapolation}. The vertical 
axis is in linear ({\bf a}) or logarithmic ({\bf b}) scale. 
The black line is a fit to the asymptotic expression 
$S(u)=c^{**}u^\delta$ of Eq.~\eqref{eq:universality}, with 
the exponent fixed to its known value $\delta=15$ and the 
fitting range $1.5\leq u\leq 1.72$. The {\bf inset} shows 
that a single amplitude $c^{**}$ describes all four models 
simultaneously, as universality requires. Error bars are one 
standard deviation.\label{fig:S}}
\end{figure}

\begin{figure*}
\centering
\includegraphics[width=\linewidth]{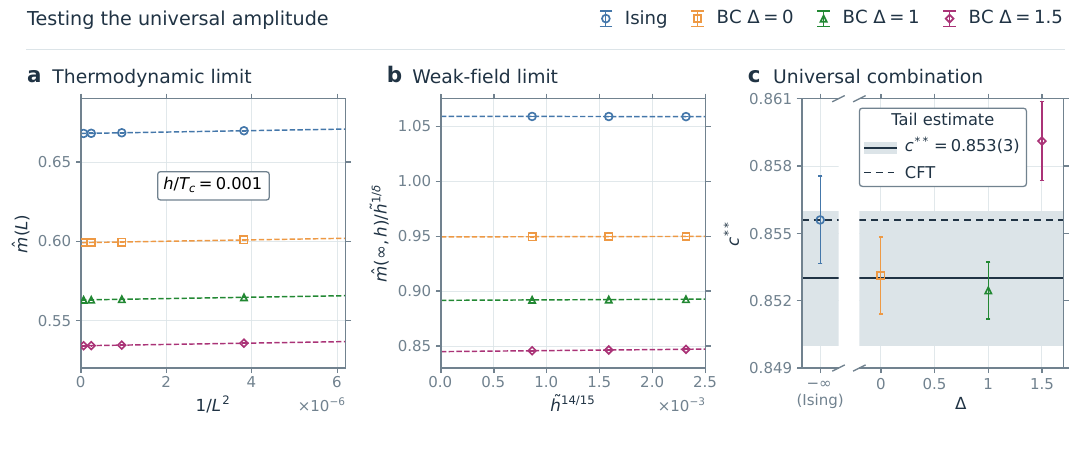}
\caption{\textbf{Numerical validation of the universal amplitude combination, Eq.~\eqref{eq:amplitudes}.} {\bf a:} 
Solution to the saddle-point equation~\eqref{eq:saddle-point}, 
giving the tethered magnetisation $\hat{m}$ at fixed external 
field, as a function of $1/L^2$ ($L$ is the linear size of 
the lattice) for $\tilde{h}=h/T_\mathrm{c}=10^{-3}$ and our 
four values of $\varDelta$. Lines are linear fits. {\bf b:} 
The rescaled magnetisation ${\langle \hat{m}\rangle_{L=\infty;
\tilde{h};\varDelta}}/{\tilde{h}^{1/\delta}}$, \emph{i.e.}, the 
l.h.s.\ of Eq.~\eqref{eq:mission:impossible}, as a function 
of $\tilde{h}^{(\delta-1)/\delta}$, $\delta=15$, for our 
four values of $\varDelta$. The lines are linear fits. 
{\bf c:} Amplitude $c^{**}$ as computed from 
Eq.~\eqref{eq:amplitudes}, shown as a function of the 
crystal field $\varDelta$ of the Blume--Capel (BC) model. The 
horizontal grey band is our estimate of $c^{**}$ from 
Eq.~\eqref{eq:final}, plus or minus one standard deviation.
The dashed horizontal line is the field-theoretical prediction, Eq.~\eqref{eq:c-CFT}.
In all three panels, error bars are one standard 
deviation.\label{fig:amplitudes}}
\end{figure*}

Our investigation of Eq.~\eqref{eq:universality} begins with 
Fig.~\ref{fig:h}, which shows $-\mathrm{d}\log\tilde{p}_L/
\mathrm{d}u$, measured directly in our TMC simulations
according to Eq.~\eqref{eq:hath}, as a function of the dimensionless magnetisation 
$u$ for our four models at several system sizes. Both scale 
invariance and model independence ---\emph{i.e.}, 
universality--- are quite evident in the central part of the 
distribution (Fig.~\ref{fig:h}--inset), where the probabilities 
are of order one. The situation is markedly different at the 
tails, as shown in the main panel, where the values of 
$-\mathrm{d}\log\tilde{p}_L/\mathrm{d}u$ differ even in order 
of magnitude across our four models (recall also 
Fig.~\ref{fig:graphical-abstract}-{\bf e}). It would be 
premature to interpret this as a universality violation, 
however, since the size dependence of 
$-\mathrm{d}\log\tilde{p}_L/\mathrm{d}u$ is significant for 
all four models. Before reaching any conclusion, one must 
extrapolate $-\mathrm{d}\log\tilde{p}_L/\mathrm{d}u$ 
separately for each model to $L\to\infty$.

We know that the leading finite-size corrections for the two-dimensional Ising 
universality class are proportional to $L^{-\gamma/\nu}$, 
$\gamma/\nu=7/4$; see, \emph{e.g.},~\cite{salas:00,bisson:25}. 
We have found good fits to our data at fixed $u$ of the form
\begin{equation}\label{eq:extrapolation}
-\frac{\mathrm{d}\log\tilde{p}_L(u)}{\mathrm{d} u}=S(u) + 
\frac{a_1(u,\varDelta)}{L^{\gamma/\nu}}+\frac{a_2(u,\varDelta)}{L^{2\gamma/\nu}}\,.
\end{equation}
The fit parameters are the infinite-volume extrapolation $S(u)$, plotted in Fig.~\ref{fig:S}, and the amplitudes $a_1$ and $a_2$. Crucially, the model ($\varDelta$) dependence disappears from our data when the thermodynamic limit is taken. To be quantitative, we have fitted our extrapolated $S(u)$ to Eq.~\eqref{eq:universality}, 
considering simultaneously the data for all four models. The only fitting parameter is the amplitude $c^{**}$. In 
the range $1.5\leq u \leq 1.72$, we obtain a figure of 
merit $\chi^2/\mathrm{dof}=66.9/91$, where 
$\mathrm{dof}=91$ denotes the number of degrees of freedom.\footnote{For $u<1.5$, the leading term alone does not adequately describe the extrapolated data.
We have simulated 
the models for $\hat{m}$ corresponding to $u>1.72$, but in 
that range the curvature of $-\mathrm{d}\log\tilde{p}_L(u)/
{\mathrm{d}u}$ is large and our interpolation, see 
Sec.~\ref{subsect:TMC}, becomes riskier, as signalled by worse 
fits to Eq.~\eqref{eq:extrapolation}.} The resulting $P$-value 
is very high, $0.97$, but this is expected given that the values of $S(u)$ at 
different $u$ for the same model are statistically correlated 
through our spline interpolation; see Sec.~\ref{subsect:TMC}. 
{To ensure a correct computation of errors, we have therefore 
obtained $c^{**}$ separately for each of the four models, 
estimating the uncertainty from the model-to-model 
fluctuations, resulting in $c^{**}=0.849(3)$. If we consider 
the subleading correction in Eq.~\eqref{eq:tail-correction}, we can extend
our fitting range to $1.2\leq u\leq 1.72$ with $\chi^2/\text{dof}=152/210$, which yields 
the final values:
\begin{align}
    c^{**} &= 0.853(3), & \prefactor&=6.640(15) . \label{eq:final}
\end{align}
The fitted value of $\psi$ lies about $5\%$ below the asymptotic prediction $(\delta-1)/2=7$. The difference exceeds the quoted uncertainty and may reflect additional corrections over the finite fitting interval.

For comparison, Malakis and Fytas~\cite{malakis2006universal} used the rms-normalised canonical magnetisation, which coincides with our scaling variable $u$ in the $L\to\infty$ limit.
Their extraction of tail parameters focused on the stable window $1.2\leq u\leq1.3$. 
}

\subsection{Universal combination of amplitudes\label{subsec:amplitudes_combination}}

To check Eq.~\eqref{eq:amplitudes} we need the two scaling 
amplitudes $A$ and $B$ for each of our models. The computation 
of $B$ is straightforward.\footnote{One separately fits for 
each crystal field $\varDelta$ to $\sqrt{\langle m^2\rangle}=
B(\varDelta)L^{-\beta/\nu}[1+b_1(\varDelta)L^{-\gamma/\nu}]$. 
The fitting parameters are the amplitudes $B(\varDelta)$ and 
$b_1(\varDelta)$; the corrections to scaling proportional to 
$L^{-\gamma/\nu}$ are specific to the two-dimensional Ising universality class.} The computation of $A(\varDelta)$, by contrast, is 
significantly more involved. Fortunately, for the Ising model 
case $\varDelta=-\infty$, we can compare our numerical approach 
to the exact result from Ref.~\cite{caselle:00b}.

The first difficulty is finding the magnetisation density in 
the thermodynamic limit for a given magnetic field $h$. We 
resolve this using ensemble equivalence~\cite{martin-mayor:11}, 
which allows thermodynamic-limit quantities to be extracted 
from finite-size TMC simulations. Specifically, we consider 
the saddle-point equation
\begin{equation}\label{eq:saddle-point}
\langle\hat{h}\rangle_{L,\hat{m}}=\tilde{h}\,
\end{equation}
where the l.h.s. is computed in our TMC simulation and $\tilde{h}$ is the external magnetic field $h$ in units of the critical temperature.
As the thermodynamic limit is approached, the 
solution of Eq.~\eqref{eq:saddle-point} differs from the
tethered expectation value $\langle\hat{m}\rangle_{L=\infty,\tilde{h}}$
by corrections of order $1/L^2$.
Unfortunately, two difficulties constrain the accessible range of $\tilde{h}$. 
On the one hand, the magnetic field cannot be too small, since 
we require $L\gg\xi$ to approach the large-$L$ limit 
($\xi\propto|\tilde{h}|^{-{1/y_h}}$ is the correlation length in 
a field). On the other hand, $\tilde{h}$ cannot be too large, 
because the strong curvature of $\langle\hat{h}\rangle_{L,\hat{m}}$ as a
function of $\hat{m}$ makes it difficult to solve 
Eq.~\eqref{eq:saddle-point}. In practice, we perform a linear 
fit in $1/L^2$ (Fig.~\ref{fig:amplitudes}-{\bf a}) to extract 
$\langle\hat{m}\rangle_{L=\infty,\tilde{h}}$ as the intercept of the fitted straight lines.
This extrapolation to $L=\infty$ has proved reliable only over the limited range 
of $\tilde{h}$ shown in Fig.~\ref{fig:amplitudes}-{\bf b}.

{Once the thermodynamic limit has been taken, we still need to 
compute the amplitude $\tilde{A}=A(1+\delta)/\delta$ through 
an extrapolation to $\tilde{h}=0$, see Eq.~\eqref{eq:A-def}. 
To this end, we note that
\begin{equation}\label{eq:mission:impossible}
\frac{\langle\hat{m}\rangle_{L=\infty,\tilde{h},\varDelta}}
{\tilde{h}^{1/\delta}}=\tilde{A}(\varDelta) + 
b_1(\varDelta)\tilde{h}^{1-\frac{1}{\delta}} + 
b_2(\varDelta)\tilde{h}^{\omega/y_h}\,,
\end{equation}}
where the $b_1$ term comes from the analytic corrections to 
scaling, the $b_2$ term accounts for the confluent corrections 
to scaling, and $\omega$ is the leading irrelevant scaling 
exponent, see, \emph{e.g.},~\cite{amit:05}. Since $\omega=2$ 
for the two-dimensional Ising universality class~\cite{caselle:02}, the 
$b_1$ term is of order $\tilde{h}^{14/15}$ while the $b_2$ 
term is of order $\tilde{h}^{16/15}$. The extrapolation to 
$\tilde{h}=0$ is therefore particularly challenging, as it 
involves two independent correction terms with very similar 
exponents. Our extrapolation procedure is illustrated in 
Fig.~\ref{fig:amplitudes}-{\bf b}. For the Ising model, our 
numerical procedure yields $\tilde{A}(\varDelta=-\infty)=
1.058\,983(9)$, in remarkable agreement with the exact result 
$1.058\,986\,12\ldots$~\cite{caselle:00b}. The accuracy of 
this extrapolation to $\tilde{h}=0$ suggests that the $b_2$ 
amplitude vanishes in the Ising-model limit 
$\varDelta\to-\infty$.

{
Our three determinations of $c^{**}$ ---from Eq.~\eqref{eq:final}, from~Eq.~\eqref{eq:amplitudes} and from CFT--- are compared in Fig.~\ref{fig:amplitudes}-\textbf{c}, showing remarkable agreement.
}

\section{Conclusions\label{sec:conclusions}}

Identifying the mechanism by which universality controls extremely rare, large fluctuations beyond the typical regime is a frontier problem in the study of critical phenomena and fills a gap in the theory of large deviations~\cite{Touchette2009}. By directly comparing four models in the two-dimensional Ising universality class at their respective critical points, we have obtained strong numerical evidence for the universality of the far tails of the probability density function of the properly normalised order parameter. Tethered Monte Carlo reconstructs this distribution down to probabilities of about $10^{-300}$, and our universality tests probe the tails down to $10^{-200}$. This extends the range accessible to earlier numerical studies by many orders of magnitude and substantially strengthens our conclusion.

Our finding of universality not only makes us optimistic about the possibility of establishing a generalised central limit theorem for critical fluctuations, but also allowed us to compute a new universal quantity, the amplitude $c^{**}$ [Eq.~\eqref{eq:universality}], in what is arguably the best-studied universality class in statistical mechanics.
{Combining the universality established here with a relation conjectured in Ref.~\cite{teza2025universal}, we argue that $c^{**}$ equals a combination of scaling amplitudes, the right-hand side of Eq.~\eqref{eq:amplitudes}, which links the free-energy amplitude in the thermodynamic limit to the finite-size-scaling amplitude of the order parameter. We provide an independent theoretical argument supporting the universality of this combination (Sec.~\ref{subsec:CFT-amplitude}): assuming that the same normalisation factor relates lattice and continuum spin fields in both regimes, all model-dependent factors cancel in the right-hand side of Eq.~\eqref{eq:amplitudes}. This yields the field-theoretical prediction $c^{**}_{\mathrm{CFT}}\simeq 0.8556$ for the square torus, which is consistent with our two numerical determinations, obtained from the tails of the probability density function and from the direct evaluation of Eq.~\eqref{eq:amplitudes}.}

{
The precision achieved here allows us to go beyond the leading stretched-exponential decay and estimate the exponent of the algebraic prefactor deep in the tails.
The independent field-theoretical prediction for $c^{**}$ provides a benchmark for the leading amplitude, and the prefactor estimate enables quantitative tests of universality at subleading order.

Several questions remain for future investigation.
It would be valuable to test whether the envisaged generalised central limit theorem extends to large fluctuations in other universality classes, such as the three-dimensional Ising or $O(N)$ models, or to other representatives of the two-dimensional Ising universality class; see, \emph{e.g.},~\cite{bruce1992}.
Extending the field-theoretical matching to other geometries and boundary conditions would allow their effects on the rms-normalised tail amplitude and on the subleading tail structure to be investigated.
}

\section{Methods\label{sec:Methods}}

\subsection{Models\label{subsect:models}}

The spins $S_{\boldsymbol{x}}\in\{0,\pm1\}$ are defined on 
the $N=L^2$ sites of a square lattice with periodic boundary 
conditions. Their interactions are described by the 
Blume--Capel Hamiltonian~\cite{blume} 
\begin{equation}\label{eq:H-def}
\mathcal{H}=\mathcal{E}-h\mathcal{M},
\end{equation}
where $h$ is the external magnetic field, 
\begin{equation}
\mathcal{M}=\sum_{\boldsymbol{x}}S_{\boldsymbol{x}}
\end{equation}
is the total magnetisation, and the interaction energy is
\begin{equation}\label{eq:E-def}
\mathcal{E}=-J\sum_{\langle\boldsymbol{x},\boldsymbol{y}
\rangle}S_{\boldsymbol{x}}S_{\boldsymbol{y}}+\varDelta
\sum_{\boldsymbol{x}}S_{\boldsymbol{x}}^{2}\,,
\end{equation}
where $\langle\boldsymbol{x},\boldsymbol{y}\rangle$ denotes 
nearest-neighbour pairs, and the crystal field $\varDelta$ 
controls the density of vacancies $(S_{\boldsymbol{x}}=0)$. 
Throughout this work we set $J=1$. In the limit 
$\varDelta\to-\infty$, vacancies are completely suppressed 
and the model reduces to the standard Ising 
model~\cite{mccoy1973two}. Varying $\varDelta$ along the critical 
line generates distinct microscopic realisations of the same 
universality class, providing the setting for the tests 
presented in this work.

For the Ising model, the inverse critical temperature is 
$\beta_{\mathrm{c}}^{\mathrm{(Ising)}}=\log(1+\sqrt{2})/2$. 
Along the continuous transition line of the Blume--Capel 
model, the critical inverse temperatures $\beta_{\mathrm{c}}
(\varDelta)$ have been determined with high numerical 
precision through finite-size scaling analyses of 
Binder-cumulant and correlation-length crossings on lattices 
with $L=4096$~\cite{bisson:25,jeon2026}.

\subsection{Tethered magnetisation and probability 
density\label{subsect:convolution}}

On a finite system containing $N=L^2$ spins, the allowed 
values of the magnetisation density $m=\mathcal{M}/N$ are 
quantised, $m_k=-1+k\Delta m$, with $k=0,1,\ldots,
\mathcal{N}$ (for the Ising model, $\Delta m=2/N$ and 
$\mathcal{N}=N$; for the Blume--Capel model, $\Delta m=1/N$ 
and $\mathcal{N}=2N$). The canonical probability of finding 
magnetisation level $m_k$ is
\begin{equation}
p_k=\frac{\sum_{\{S_{\boldsymbol{x}}\}}\exp(-\beta\mathcal{E})
\,\delta_{\mathcal{M},m_k N}}{\sum_{\{S_{\boldsymbol{x}}\}}
\exp(-\beta\mathcal{E})}\,,
\end{equation}
where $\delta_{a,b}$ stands for Kronecker's delta symbol. 
The finite-system magnetisation density is then
\begin{equation}\label{eq:rho-naive}
\rho_L(m)=\sum_{k=0}^{\mathcal{N}}p_k\,\delta(m-m_k)\,,
\end{equation}
where $\delta(m-m_k)$ denotes Dirac's delta distribution. 
Since all simulations are performed at criticality, we 
suppress the dependence of $\rho(m)$ on $\beta$ throughout.  

At finite $L$, the distribution $\rho_L(m)$ in Eq.~\eqref{eq:rho-naive} is a comb of Dirac delta functions.
To work with a smooth probability density at finite size, we introduce a continuously distributed random variable $\hat{m}$ through convolution with a Gaussian filter:
\begin{equation}\label{eq:Omega-def}
p_L(\hat{m})=
\int_{-\infty}^{\infty}\dd m\,\rho_L(m)\,
\sqrt{\frac{N}{2\pi}}\,
\ee^{-N(\hat{m}-m)^2/2}
=
\ee^{-N\varOmega_L(\hat{m})}\,,
\end{equation}
where the last equality defines the Helmholtz potential 
$\varOmega_L(\hat{m})$. The Gaussian convolution effectively 
averages over $\sqrt{N}=L^{d/2}$ of the discrete 
magnetisation levels $m_k$, implying a resolution scale 
$\Delta\hat{m}\sim N^{-1/2}=L^{-d/2}$. This loss of 
resolution is negligible because the critical fluctuations 
of the magnetisation occur at the much larger scale 
$m\sim L^{-\beta/\nu}$; in the two-dimensional Ising 
universality class, $L^{-\beta/\nu}=L^{-1/8}\gg L^{-1}$. 
More generally, the same situation arises whenever the 
magnetic susceptibility $\chi$ diverges at the critical 
point in the thermodynamic limit, since $\chi\propto 
L^{d-2\beta/\nu}$ if hyperscaling holds; see, 
\emph{e.g.},~\cite{amit:05}.

The physical equivalence of $m$ and $\hat{m}$ can also be 
illustrated by comparing the generating functions for 
$\rho_L(m)$ and $p_L(\hat{m})$. Setting $\tilde{h}=h/T_\mathrm{c}$,
\begin{eqnarray}
\int_{-\infty}^\infty\mathrm{d}m\,\ee^{N\tilde{h}m}\,
\rho_L(m)&=&\ee^{N[g_L(\tilde{h})-g_L(0)]}\,,
\label{eq:gL-def}\\
\int_{-\infty}^\infty\mathrm{d}\hat{m}\,
\ee^{N\tilde{h}\hat{m}}\,p_L(\hat{m})&=&
\ee^{N\hat{g}_L(\tilde{h})}\,,\\
\hat{g}_L(\tilde{h})&=&\frac{\tilde{h}^2}{2}+
g_L(\tilde{h})-g_L(0)\,,\label{eq:hat-g-def}
\end{eqnarray}
where Eq.~\eqref{eq:gL-def} defines the finite-volume 
free-energy density $g_L(\tilde{h})$.\footnote{We do not 
include the temperature dependence in $g_L$ because we are 
working at fixed temperature $T=T_\mathrm{c}$.} It is well 
known~\cite{parisi:88} that $g_L$ converges as $L\to\infty$ 
to the free-energy density $g(\tilde{h})$. 
{Equation~\eqref{eq:hat-g-def} implies that $\hat{g}_L(\tilde{h})$ also has a thermodynamic limit.
The leading field-dependent singularity is
\begin{equation}\label{eq:sing}
g(\tilde{h})-g(0)\sim A|\tilde{h}|^{d/y_h}.
\end{equation}
Since $d/y_h=1+1/\delta<2$, the additional term $\tilde{h}^2/2$ in $\hat{g}$ is subleading.
Differentiating Eq.~\eqref{eq:hat-g-def} gives
\begin{equation}
\langle\hat{m}\rangle_{L=\infty,\tilde{h}}
=
\langle m\rangle_{L=\infty,\tilde{h}}+\tilde{h}.
\end{equation}
The linear shift is subleading to $\tilde{h}^{1/\delta}$, so Eq.~\eqref{eq:A-def} also describes the leading response of $\langle\hat{m}\rangle$, with the same amplitude $A(\delta+1)/\delta$.

At zero external field, all cumulants of $m$ and $\hat{m}$ coincide except the second-order cumulant.
The Gaussian convolution gives
\begin{equation}
\frac{\langle\hat{m}^2\rangle_L}{\langle m^2\rangle_L}-1
=
\frac{1}{N\langle m^2\rangle_L}
\propto L^{-(d-2\beta/\nu)}.
\end{equation}
}

Turning now to the dimensionless variable introduced in 
Sec.~\ref{sec:Intro}, rather than working with $\hat{m}$ 
directly, it is convenient to define
\begin{equation}\label{eq:u-def}
u=\hat{m}/\sqrt{\langle m^2\rangle_L}\,,
\end{equation}
where $\langle\cdots\rangle$ denotes the canonical expectation 
value at the critical inverse temperature $\beta_{\mathrm{c}}
(\varDelta)$. The corresponding probability density is
\begin{equation}\label{eq:change-to-u}
\tilde{p}_L(u)=\sqrt{\langle m^2\rangle_L}\,~p_L\!\left(\hat{m}=
u\sqrt{\langle m^2\rangle_L}\right).
\end{equation}
Our primary objective is to establish pointwise convergence, 
as $L\to\infty$, towards a universal limiting distribution. 
For sufficiently large $L$, we may replace the normalising 
factor $\sqrt{\langle m^2\rangle_L}$ by $BL^{-\beta/\nu}$, 
see Eq.~\eqref{eq:B-def}. Using the definition of 
$p_L(\hat{m})$ in Eq.~\eqref{eq:Omega-def} and substituting 
$z=mL^{\beta/\nu}$ in Eq.~\eqref{eq:change-to-u}, we find
\begin{eqnarray}
\tilde{p}_L(u)&\approx& B\int_{-\infty}^{\infty}\mathrm{d}z\,
\sqrt{\frac{L^{d-\frac{2\beta}{\nu}}}{2\pi}}\,
\ee^{-\frac{1}{2}L^{d-\frac{2\beta}{\nu}}(uB-z)^2}
\times\nonumber\\
&\times&L^{-\beta/\nu}\rho_L(m=zL^{-\beta/\nu})\,.
\label{eq:delta-approximant}
\end{eqnarray}
Assuming, as is customary in finite-size scaling studies, 
that the limit
\begin{equation}\label{eq:FSS-assumption}
\lim_{L\to\infty}L^{-\beta/\nu}\rho_L(m=zL^{-\beta/\nu})
=f_1(z)
\end{equation}
exists and is a smooth function $f_1(z)$,\footnote{At finite $L$, the expression under the limit in Eq.~\eqref{eq:FSS-assumption} is a sum of Dirac masses. The quantity that converges 
to a limit as $L$ grows is the cumulative distribution 
function, rather than the probability density.} 
Eq.~\eqref{eq:delta-approximant} gives
\begin{equation}\label{eq:final-limit}
\lim_{L\to\infty}\tilde{p}_L(u)=Bf_1(z=uB)\,,
\end{equation}
whenever $d-2\beta/\nu>0$.

In practice, rather than $\tilde{p}_L(u)$ itself, we compute 
its logarithmic derivative, which is related to the Helmholtz 
potential through
\begin{equation}\label{eq:def-u-derivative}
-\frac{\mathrm{d}\log\tilde{p}_L}{\mathrm{d}u}=
N\sqrt{\langle m^2\rangle_L}~\,\partial_{\hat{m}}\varOmega_L
\!\left(\hat{m}=u\sqrt{\langle m^2\rangle_L}~\right).
\end{equation}
The remaining task is therefore to determine the derivative 
of the Helmholtz potential ---precisely the quantity that 
Tethered Monte Carlo gives direct access to, as described 
in Sec.~\ref{subsect:TMC}.

\subsection{Finite-size scaling, thermodynamic limits
and critical singularities\label{subsec:FSS}}

Consider $z=L^{\beta/\nu}m$ at the critical point.
It was conjectured~\cite{bouchaud:90,Bruce1995,Hilfer1995} that its limiting density $f_1(z)$, Eq.~\eqref{eq:FSS-assumption}, has the leading logarithmic asymptotics
\begin{equation}\label{eq:assymptotic-1}
\log f_1(z)\sim -c|z|^{1+\delta}
\qquad (|z|\to\infty).
\end{equation}
Rigorous results for the corresponding logarithmic magnetisation-tail asymptotics in the critical two-dimensional Ising model are available in Refs.~\cite{camia2016planar,camia2025power}.
More recently, it was shown~\cite{teza2025universal} that this leading behaviour can be derived from a simple hypothesis, which can be directly verified for a mean-field model and for an anomalous-diffusion model.
Specifically,
we assume that, for a properly normalised magnetic field, 
the cumulant-generating function for $f_1(z)$,
\begin{equation}
\bar{G}(L^{y_h}\tilde{h})=\log\left[\int_{-\infty}^{+\infty}
\dd z\,\ee^{zL^{y_h}\tilde{h}}f_1(z)\right],
\end{equation}
is extensive (\emph{i.e.}, it grows as $L^d$ in the 
thermodynamic limit). Note the mixed nature of $\bar{G}$: 
although it is defined in a finite volume, it contains the 
limiting distribution $f_1(z)$. Equation~\eqref{eq:assymptotic-1} 
then follows from the hypothesis of linear scaling with the
number of spins for $\bar{G}$, with the additional result 
that the scaling amplitude $c$ is 
predicted~\cite{teza2025universal} to be simply related to 
the amplitude $A$ defined in Eq.~\eqref{eq:A-def}:
\begin{equation}\label{eq:amplitudes-A-and-c}
A=\frac{\delta}{c^{1/\delta}}
\left(\frac{1}{\delta+1}\right)^{(1+\delta)/\delta}\,.
\end{equation}
{We recall that $A$ also governs the leading singular behaviour of the free-energy density as $\tilde{h}\to0$ at $T_{\mathrm{c}}$, according to Eq.~\eqref{eq:sing}.}

Equation~\eqref{eq:amplitudes-A-and-c} tells us that 
amplitude $c$ is just as universal as the free-energy 
density (and its associated amplitude $A$). In particular, 
changes of boundary conditions and/or domain geometry, 
which do affect the central part of the $f_1(z)$ 
distribution~\cite{kaneda2001finite}, leave amplitude $c$ unvaried, {consistently also with the particular results in Ref.\cite{camia2025power}}. Finally, 
expressing the thermodynamic limit in terms of 
$u=\hat{m}/\sqrt{\langle m^2\rangle}$, as we do throughout 
this work, the combination of Eq.~\eqref{eq:final-limit} 
and Eq.~\eqref{eq:amplitudes-A-and-c} gives
\begin{equation}\label{eq:assymptotic-2}
\log\tilde{p}(u)\sim
-\frac{c^{**}|u|^{1+\delta}}{\delta+1}
\qquad (|u|\to\infty),
\end{equation}
where $\tilde{p}(u)=\lim_{L\to\infty}\tilde{p}_L(u)$ and $c^{**}$ is given in Eq.~\eqref{eq:amplitudes}.

Two important questions remain open at this point: (i) is 
$\bar{G}$ truly an extensive quantity, {growing as the leading singular part of $L^dg_L(\tilde{h})$ for $L$ approaching $\infty$?}, and (ii) is 
amplitude $c^{**}$ truly universal? We do not have a fully 
satisfactory answer to either of these questions, but we 
can offer a plausibility argument that provides positive 
answers to both.

Finite-size scaling, see for instance~\cite{barber:83}, 
predicts for the free-energy density at the critical 
temperature
\begin{equation}\label{eq:FSS-barber}
L^d[g_L(\tilde{h})-g_{\mathrm{reg}}(\tilde{h})]\simeq
\varPhi(\tilde{h}L^{y_h}/h_0,u_\omega L^{-\omega},\ldots)\,,
\end{equation}
where $g_{\mathrm{reg}}$ is the analytic background and $\varPhi$ is the universal scaling function.
Here $\omega$
is the leading irrelevant scaling exponent at the 
renormalisation-group fixed point, the dots stand for the 
contribution of subleading irrelevant scaling fields, and 
$h_0$ and $u_\omega$ are system-dependent constants related 
to the scaling fields; see, \emph{e.g.},~\cite{amit:05}.
\footnote{Note that $\varPhi$ is universal with respect to 
changes in the Hamiltonian, but it varies with boundary 
conditions and/or domain geometry.}
By Eq.~\eqref{eq:gL-def}, $L^d[g_L(\tilde{h})-g_L(0)]$ is the finite-volume cumulant-generating function of $z$ at $s=L^{y_h}\tilde{h}$.
The $r$th cumulant $\mu_{r,z}$ of $z=L^{\beta/\nu}m=L^{d-y_h}m$ is given by
\begin{equation}
\mu_{r,z}=\left.\frac{1}{L^{ry_h}}\frac{\partial^r}
{\partial\tilde{h}^r}[L^d g_L(\tilde{h})]\right|_{\tilde{h}=0}\,.
\end{equation}
At fixed $s$, the even analytic background contributes $L^d[g_{\mathrm{reg}}(\tilde{h})-g_{\mathrm{reg}}(0)]=O(L^{d-2y_h})$, which vanishes since $2y_h>d$.
Assuming convergence of the finite-volume generator to that of $f_1(z)$, Eq.~\eqref{eq:FSS-barber} gives $\bar{G}(s)=\varPhi(s/h_0,0,\ldots)-\varPhi(0,0,\ldots)$.
Matching its large-$|s|$ behaviour to $L^d$ times the leading singular term in Eq.~\eqref{eq:sing} motivates $\bar{G}(s)\sim A|s|^{d/y_h}$ and hence the assumed extensivity at fixed $\tilde{h}$.

As for the universality of $c^{**}$, the cumulants of $u$ 
(all cumulants of odd order, $r=2k+1$, vanish) satisfy
\begin{equation}
\lim_{L\to\infty}\mu_{2k,u}=\frac{\varPhi^{(2k,0,\ldots)}
(0,0,\ldots)}{[\varPhi^{(2,0,\ldots)}(0,0,\ldots)]^k}\,,
\end{equation}
where $\varPhi^{(k,0,\ldots)}$ indicates the $k$th 
derivative with respect to the first argument of $\varPhi$. 
This is the standard argument for the universality of the 
large-$L$ limit of the cumulants of $u$. Under the 
additional assumption that the limits of the cumulants 
coincide with the cumulants of the limiting distribution 
$Bf_1(z=uB)$, we conclude that $Bf_1(z=uB)$ is universal 
---including its tails described by 
Eq.~\eqref{eq:assymptotic-2}\footnote{Given the fast decay 
at large values of $|u|$, Eq.~\eqref{eq:assymptotic-2}, 
the cumulants of $u$ determine the probability 
distribution~\cite{chihara:78}.}--- with the standard 
caveats regarding boundary conditions and domain geometry.

\subsection{Field-theoretical prediction of the tail amplitude\label{subsec:CFT-amplitude}}

Using the magnetisation amplitude $\tilde{A}=(1+\delta)A/\delta$, already introduced in the Results, Eq.~\eqref{eq:amplitudes} becomes, for $\delta=15$,
\begin{equation}
c^{**}=\frac{B^{16}}{\tilde{A}^{15}}.
\end{equation}
The continuum counterparts of these two amplitudes are denoted by $B_{\mathrm{CFT}}$ and $\tilde{A}_{\mathrm{CFT}}$.
Both are evaluated using the continuum spin-field normalisation adopted in Refs.~\cite{salas:00,caselle:00b}.
In this convention, the critical zero-field two-point spin correlation on the infinite plane equals $r^{-1/4}$, where $r$ is the spatial separation between the two spins  (on the torus, the correlation function is more complex~\cite{difrancesco:87,difrancesco:88}).

For the square torus, $B_{\mathrm{CFT}}^2$ is the integrated two-point function on the unit square computed in Eq.~(2.12) of Ref.~\cite{salas:00}:
\begin{equation}
B_{\mathrm{CFT}}^2=1.552\,432\,954\,65(4).
\end{equation}
The bulk critical magnetisation amplitude is
\begin{equation}
\tilde{A}_{\mathrm{CFT}}=1.277\,582\,27\ldots,
\end{equation}
which is denoted $A_M$ in Table~1 of Ref.~\cite{caselle:00b}.

Let $R$ be the model-dependent factor relating the microscopic spins $S_{\boldsymbol{x}}$ to the continuum spin field at leading scaling order. We assume that this same factor $R$ applies both to the infinite plane (associated with $\tilde{A}$) and to the finite-size torus at criticality (associated with $B$).

The lattice amplitudes then satisfy
\begin{equation}
B=R B_{\mathrm{CFT}},
\qquad
\tilde{A}=R^{16/15}\tilde{A}_{\mathrm{CFT}}.
\end{equation}
The second relation is Eq.~(57) of Ref.~\cite{caselle:00b}, expressed in our notation.
Its exponent accounts for one factor $R$ from the magnetisation and a factor $R^{1/15}$ from the rescaling of its conjugate field.

For the square-lattice Ising model, the matching factor inferred from the known bulk amplitudes~\cite{caselle:00b} predicts $B=1.044\,964\,37\ldots$.
This agrees with our independently measured value $B=1.044\,97(11)$, providing a numerical consistency check of the matching assumption.

Combining these expressions gives
\begin{equation}
\begin{aligned}
c^{**}_{\mathrm{CFT}}
&=\frac{(R B_{\mathrm{CFT}})^{16}}
{\bigl[R^{16/15}\tilde{A}_{\mathrm{CFT}}\bigr]^{15}}=\frac{B_{\mathrm{CFT}}^{16}}
{\tilde{A}_{\mathrm{CFT}}^{15}}
\approx0.855\,598.
\end{aligned}
\end{equation}
The model-dependent normalisation factor $R$ cancels.
The dependence on macroscopic geometry, topology and boundary conditions remains encoded in $B_{\mathrm{CFT}}$.
The quoted numerical prediction applies specifically to a square torus, corresponding to a square domain with periodic boundary conditions.

\begin{figure}[ht!]
\centering
\includegraphics[width=\linewidth]{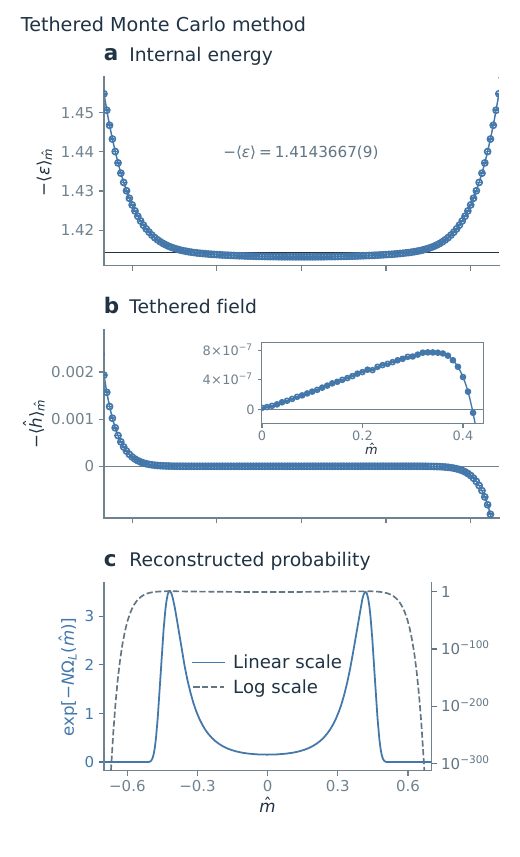}
\caption{\textbf{The Tethered Monte Carlo method illustrated 
for a $\boldsymbol{4096\times 4096}$ Ising system.} 
\textbf{a.}~Fixed-$\hat{m}$ simulations are performed across 
a mesh of tethered parameter values, from which tethered 
expectation values of all observables of interest are 
measured. 
\textbf{b.}~The key observable is $\langle\hat{h}\rangle_
{\hat{m}}$, which provides direct access to the derivative 
of the Helmholtz potential and hence to the probability 
density $p_L(\hat{m})$. 
\textbf{c.}~Numerical integration of $\langle\hat{h}\rangle_
{\hat{m}}$ reconstructs $p_L(\hat{m})$ through 
Eq.~\eqref{eq:hath} ---equivalently, 
Eq.~\eqref{eq:TMC2}--- enabling the probability density 
to be determined down to probabilities of approximately 
$10^{-300}$ (note the logarithmic scale), while maintaining 
high accuracy for the corresponding canonical observables. 
Canonical expectation values are then obtained by integrating 
tethered observables over $\hat{m}$ with weight 
$p_L(\hat{m})$, see Eq.~\eqref{eq:canonical}. As a 
validation, the reconstructed energy density, $\langle
\epsilon\rangle=-1.414\,366\,7(9)$, is in excellent 
agreement with the exact result, $\langle\epsilon\rangle^
{\mathrm{(exact)}}=-1.414\,365\,525\ldots$, obtained from the formulas of Ref.~\cite{ferdinand:69}.
\label{fig:TMC}}
\end{figure}

\subsection{Tethered Monte Carlo: Implementation \label{subsect:TMC}}

Following Ref.~\cite{fernandez:09}, we introduce the tethered 
ensemble (employing the ``linear'' variant described in 
Refs.~\cite{martin-mayor:11,yllanes:11}), in which the 
externally fixed parameters are the inverse temperature 
$\beta$ and the tethered magnetisation $\hat{m}$. The 
expectation value of an observable $O(\{S_{\boldsymbol{x}}\})$ 
is defined through
\begin{align}\label{eq:TMC}
\omega(\hat{m};\{S_{\boldsymbol{x}}\})&\propto
\ee^{-\beta\mathcal{E}-N(\hat{m}-m)^2/2}\,,\\[1mm]
\langle O\rangle_{\hat{m}}&=
\frac{\sum_{\{S_{\boldsymbol{x}}\}}O(\{S_{\boldsymbol{x}}\})
\,\omega(\hat{m};\{S_{\boldsymbol{x}}\})}
{\sum_{\{S_{\boldsymbol{x}}\}}\omega(\hat{m};
\{S_{\boldsymbol{x}}\})}\,.
\end{align}
One can show that the derivative of the Helmholtz potential, 
Eq.~\eqref{eq:Omega-def}, is given by the tethered 
expectation value
\begin{equation}\label{eq:TMC2}
\frac{\partial\varOmega_L}{\partial\hat{m}}=
\langle\hat{h}\rangle_{\hat{m}}\,,\qquad
\hat{h}(\{S_{\boldsymbol{x}}\})=
\hat{m}-m(\{S_{\boldsymbol{x}}\})\,.
\end{equation}
Equation~\eqref{eq:TMC2} determines $\varOmega_L(\hat{m})$ 
up to an additive constant, which is fixed by the 
normalisation condition
\begin{equation}
1=\int\dd\hat{m}\,p_L(\hat{m})=
\int\dd\hat{m}\,\ee^{-N\varOmega_L(\hat{m})}\,.
\end{equation}
Canonical expectation values are then obtained by integrating 
tethered expectation values over $\hat{m}$, weighted by 
$p_L(\hat{m})$:
\begin{equation}\label{eq:canonical}
\langle O\rangle=\int\dd\hat{m}\,p_L(\hat{m})\,
\langle O\rangle_{\hat{m}}=
\int\dd\hat{m}\,\ee^{-N\varOmega_L(\hat{m})}
\langle O\rangle_{\hat{m}}\,.
\end{equation}

The tethered weight in Eq.~\eqref{eq:TMC} is sampled with 
a standard single-spin Metropolis algorithm; see, 
\emph{e.g.}, Ref.~\cite{sokal:97}. At each lattice site 
$\boldsymbol{x}$, a trial value $S_{\boldsymbol{x}}^
{\mathrm{new}}\neq S_{\boldsymbol{x}}^{\mathrm{old}}$ is 
selected uniformly from the two remaining spin states 
(\emph{i.e.}, the two values in $\{0,\pm 1\}$ other than 
$S_{\boldsymbol{x}}^{\mathrm{old}}$) and accepted with 
probability $\min\left\{1,\frac{\omega^{\mathrm{new}}}
{\omega^{\mathrm{old}}}\right\}$. Sites are updated sequentially in lexicographic order.

Although the local Metropolis algorithm satisfies all the 
requirements for a correct Monte Carlo 
simulation~\cite{sokal:97}, equilibration can be 
significantly accelerated by supplementing it with a cluster 
update. In our implementation, the cluster algorithm acts 
only on the subgraph formed by sites with 
$S_{\boldsymbol{x}}=\pm 1$. The update allows these spins 
to change sign, while transitions of the form 
$S_{\boldsymbol{x}}\to 0$ are forbidden. Subject to this 
restriction, the algorithm is identical to that of 
Ref.~\cite{martin-mayor:09}, to which we refer the reader 
for a detailed description; the only minor difference lies 
in the precise definition of the weight $\omega$ in 
Eq.~\eqref{eq:TMC}. Because the cluster update conserves 
$S_{\boldsymbol{x}}^2$ at every site, it is not ergodic by 
itself and must be combined with local Metropolis updates 
to obtain a correct Monte Carlo sampling of the weight in 
Eq.~\eqref{eq:TMC}. Our simulations interleave one 
full-lattice Metropolis sweep with a cluster update. 
Once the clusters are drawn, we must keep in mind that, unlike 
in a canonical simulation, the different cluster orientations
are not equiprobable, because they do not conserve the total magnetisation. As explained in~\cite{martin-mayor:09}, we address this by selecting only $5$ clusters with probability proportional to their sizes and flipping them according to the tethered weight. We repeat this process for a number $N_\text{flip}$ of iterations that grows linearly in $L^2$, reaching $N_\text{flip}=2^{19}$ for 
$L=4096$.

The overall TMC procedure consists of the following steps:
\begin{enumerate}
\item Select a mesh of tethered parameters 
$\{\hat{m}_s\}_{s=1}^{M}$; see Ref.~\cite{fernandez:09} 
for guidance on the choice of mesh. Here we use equally spaced $\hat m_s$, with $\Delta \hat m=0.01$ for Ising and $\Delta \hat m=0.009$ for Blume-Capel (taking into account that $\langle m^2\rangle$ is larger for the Ising case).
\item Perform an independent TMC simulation at each value 
of $\hat{m}_s$, measuring $\langle\hat{h}\rangle_{\hat{m}_s}$ 
together with the tethered expectation values 
$\langle O\rangle_{\hat{m}_s}$ of the observables of 
interest. We run for $10^6$ Metropolis + cluster updates 
for each $\hat m_s$.
\item Reconstruct the Helmholtz potential 
$\varOmega_L(\hat{m})$ by numerically integrating 
$\langle\hat{h}\rangle_{\hat{m}}$ over the mesh; see 
Ref.~\cite{fernandez:09} for details of the integration 
procedure. We use cubic splines to generate smooth $\langle O\rangle_{\hat{m}}$ curves
from our $\hat m$ grid.
\item Optionally, recover canonical expectation values 
$\langle O\rangle$ by averaging $\langle O\rangle_{\hat{m}}$ 
over $\hat{m}$, weighted by the canonical probability 
$\exp[-N\varOmega_L(\hat{m})]$, see Eq.~\eqref{eq:canonical}.
\end{enumerate}
Figure~\ref{fig:TMC} illustrates the procedure for the 
internal energy $\epsilon$, demonstrating that canonical 
expectation values computed via TMC are in excellent 
agreement with exact results. We have also checked 
that the universal amplitude ratios ---the correlation length 
in units of the system size and the Binder cumulant---
computed in our simulations agree with their values
obtained with CFT.

Comparing results across models at fixed $u=\hat{m}/
\sqrt{\langle{m}^2\rangle}$ requires interpolating 
between the simulated values on our $\hat{m}$ grid.
The main source of statistical error in this process
is the uncertainty $\sigma_{\langle m^2\rangle}$ of the scaling factor $\langle m^2\rangle$, 
much larger than that of the individual $\langle O\rangle_{\hat{m}_s}$. 
We therefore estimate the error bar of an interpolated average as the difference between the values obtained by rescaling with $\langle  m^2\rangle+\sigma_{\langle m^2\rangle}$
and with $\langle m^2\rangle-\sigma_{\langle m^2\rangle}$, when determining the value of $\hat m$ 
that corresponds to the target value of $u$. For 
our key observable, $\langle\hat{h}\rangle_{\hat{m}}$, 
this interpolation is reliable up to probabilities around 
$\tilde{p}_L(u)\sim 10^{-200}$, corresponding to $u\approx 1.76$
for our largest lattices. Beyond this point, the derivative 
of $\langle\hat{h}\rangle_{\hat{m}}$ grows rapidly (see 
Fig.~\ref{fig:TMC}-{\bf b}), meaning that a finer mesh 
spacing $\Delta\hat{m}$ would be required to avoid 
systematic errors at our level of statistical precision. 
For this reason, and to ensure the absence of systematic 
effects, we have restricted our finite-size scaling 
analysis to $u\leq 1.72$.

\begin{acknowledgments}
G.T. acknowledges support by the Max Planck Society.
The work of N.G. Fytas and A. Vasilopoulos was supported by the 
Leverhulme Trust through a Research Project Grant 
(RPG-2026-025). This work was partially supported by Ministerio de Ciencia, Innovación y Universidades (Spain) and by the European Regional Development Fund through grants no. PID2022-136374NB-C21 and PID2024-158623NB-C22 (MCIU/AEI/10.13039/501100011033/FEDER, UE).
\end{acknowledgments}

\bibliography{refs.bib}

\providecommand{\noopsort}[1]{}\providecommand{\singleletter}[1]{#1}%
\begin{thebibliography}{48}%
\makeatletter
\providecommand \@ifxundefined [1]{%
 \@ifx{#1\undefined}
}%
\providecommand \@ifnum [1]{%
 \ifnum #1\expandafter \@firstoftwo
 \else \expandafter \@secondoftwo
 \fi
}%
\providecommand \@ifx [1]{%
 \ifx #1\expandafter \@firstoftwo
 \else \expandafter \@secondoftwo
 \fi
}%
\providecommand \natexlab [1]{#1}%
\providecommand \enquote  [1]{``#1''}%
\providecommand \bibnamefont  [1]{#1}%
\providecommand \bibfnamefont [1]{#1}%
\providecommand \citenamefont [1]{#1}%
\providecommand \href@noop [0]{\@secondoftwo}%
\providecommand \href [0]{\begingroup \@sanitize@url \@href}%
\providecommand \@href[1]{\@@startlink{#1}\@@href}%
\providecommand \@@href[1]{\endgroup#1\@@endlink}%
\providecommand \@sanitize@url [0]{\catcode `\\12\catcode `\$12\catcode
  `\&12\catcode `\#12\catcode `\^12\catcode `\_12\catcode `\%12\relax}%
\providecommand \@@startlink[1]{}%
\providecommand \@@endlink[0]{}%
\providecommand \url  [0]{\begingroup\@sanitize@url \@url }%
\providecommand \@url [1]{\endgroup\@href {#1}{\urlprefix }}%
\providecommand \urlprefix  [0]{URL }%
\providecommand \Eprint [0]{\href }%
\providecommand \doibase [0]{https://doi.org/}%
\providecommand \selectlanguage [0]{\@gobble}%
\providecommand \bibinfo  [0]{\@secondoftwo}%
\providecommand \bibfield  [0]{\@secondoftwo}%
\providecommand \translation [1]{[#1]}%
\providecommand \BibitemOpen [0]{}%
\providecommand \bibitemStop [0]{}%
\providecommand \bibitemNoStop [0]{.\EOS\space}%
\providecommand \EOS [0]{\spacefactor3000\relax}%
\providecommand \BibitemShut  [1]{\csname bibitem#1\endcsname}%
\let\auto@bib@innerbib\@empty
%</preamble>
\bibitem [{\citenamefont {Botet}\ and\ \citenamefont
  {P{\l}oszajczak}(2002)}]{alma991005852086808044}%
  \BibitemOpen
  \bibfield  {author} {\bibinfo {author} {\bibfnamefont {R.}~\bibnamefont
  {Botet}}\ and\ \bibinfo {author} {\bibfnamefont {M.}~\bibnamefont
  {P{\l}oszajczak}},\ }\href@noop {} {\emph {\bibinfo {title} {Universal
  Fluctuations: The Phenomenology of Hadronic Matter}}},\ \bibinfo {series}
  {World Scientific Lecture Notes in Physics}, Vol.~\bibinfo {volume} {65}\
  (\bibinfo  {publisher} {World Scientific Publishing Co.},\ \bibinfo {address}
  {Singapore},\ \bibinfo {year} {2002})\BibitemShut {NoStop}%
\bibitem [{\citenamefont {Kadanoff}(2000)}]{kadanoff2000statistical}%
  \BibitemOpen
  \bibfield  {author} {\bibinfo {author} {\bibfnamefont {L.~P.}\ \bibnamefont
  {Kadanoff}},\ }\href {https://doi.org/10.1142/4016} {\emph {\bibinfo {title}
  {Statistical Physics: Statics, Dynamics and Renormalization}}}\ (\bibinfo
  {publisher} {World Scientific},\ \bibinfo {address} {Singapore},\ \bibinfo
  {year} {2000})\BibitemShut {NoStop}%
\bibitem [{\citenamefont {Cassandro}\ and\ \citenamefont
  {Jona-Lasinio}(1978)}]{cassandro1978critical}%
  \BibitemOpen
  \bibfield  {author} {\bibinfo {author} {\bibfnamefont {M.}~\bibnamefont
  {Cassandro}}\ and\ \bibinfo {author} {\bibfnamefont {G.}~\bibnamefont
  {Jona-Lasinio}},\ }\bibfield  {title} {\bibinfo {title} {Critical point
  behaviour and probability theory},\ }\href
  {https://doi.org/10.1080/00018737800101504} {\bibfield  {journal} {\bibinfo
  {journal} {Adv. Phys.}\ }\textbf {\bibinfo {volume} {27}},\ \bibinfo {pages}
  {913} (\bibinfo {year} {1978})}\BibitemShut {NoStop}%
\bibitem [{\citenamefont {Jona-Lasinio}(2001)}]{jona2001renormalization}%
  \BibitemOpen
  \bibfield  {author} {\bibinfo {author} {\bibfnamefont {G.}~\bibnamefont
  {Jona-Lasinio}},\ }\bibfield  {title} {\bibinfo {title} {Renormalization
  group and probability theory},\ }\href
  {https://doi.org/10.1016/S0370-1573(01)00042-4} {\bibfield  {journal}
  {\bibinfo  {journal} {Phys, Rep.}\ }\textbf {\bibinfo {volume} {352}},\
  \bibinfo {pages} {439} (\bibinfo {year} {2001})}\BibitemShut {NoStop}%
\bibitem [{\citenamefont {Teza}\ and\ \citenamefont
  {Stella}(2025)}]{teza2025universal}%
  \BibitemOpen
  \bibfield  {author} {\bibinfo {author} {\bibfnamefont {G.}~\bibnamefont
  {Teza}}\ and\ \bibinfo {author} {\bibfnamefont {A.~L.}\ \bibnamefont
  {Stella}},\ }\bibfield  {title} {\bibinfo {title} {{Universal and
  Nonuniversal Signatures in the Scaling Functions of Critical Variables}},\
  }\href {https://doi.org/10.1103/PhysRevLett.134.127102} {\bibfield  {journal}
  {\bibinfo  {journal} {Phys. Rev. Lett.}\ }\textbf {\bibinfo {volume} {134}},\
  \bibinfo {pages} {127102} (\bibinfo {year} {2025})}\BibitemShut {NoStop}%
\bibitem [{\citenamefont {Binder}(1981)}]{binder1981}%
  \BibitemOpen
  \bibfield  {author} {\bibinfo {author} {\bibfnamefont {K.}~\bibnamefont
  {Binder}},\ }\bibfield  {title} {\bibinfo {title} {{Finite size scaling
  analysis of Ising model block distribution functions}},\ }\href
  {https://doi.org/10.1007/BF01293604} {\bibfield  {journal} {\bibinfo
  {journal} {Z. Phys. B}\ }\textbf {\bibinfo {volume} {43}},\ \bibinfo {pages}
  {119} (\bibinfo {year} {1981})}\BibitemShut {NoStop}%
\bibitem [{\citenamefont {Gallavotti}\ and\ \citenamefont
  {Martin-L{\"o}f}(1975)}]{gallavotti1975}%
  \BibitemOpen
  \bibfield  {author} {\bibinfo {author} {\bibfnamefont {G.}~\bibnamefont
  {Gallavotti}}\ and\ \bibinfo {author} {\bibfnamefont {A.}~\bibnamefont
  {Martin-L{\"o}f}},\ }\bibfield  {title} {\bibinfo {title} {Block-spin
  distributions for short-range attractive ising models},\ }\href
  {https://doi.org/10.1007/BF02737692} {\bibfield  {journal} {\bibinfo
  {journal} {Il Nuovo Cimento B (1971-1996)}\ }\textbf {\bibinfo {volume}
  {25}},\ \bibinfo {pages} {425} (\bibinfo {year} {1975})}\BibitemShut
  {NoStop}%
\bibitem [{\citenamefont {Bruce}(1995)}]{Bruce1995}%
  \BibitemOpen
  \bibfield  {author} {\bibinfo {author} {\bibfnamefont {A.}~\bibnamefont
  {Bruce}},\ }\bibfield  {title} {\bibinfo {title} {{Critical finite-size
  scaling of the free energy}},\ }\href
  {https://doi.org/10.1088/0305-4470/28/12/008} {\bibfield  {journal} {\bibinfo
   {journal} {J Phys. A: Math. Gen.}\ }\textbf {\bibinfo {volume} {28}},\
  \bibinfo {pages} {3345} (\bibinfo {year} {1995})}\BibitemShut {NoStop}%
\bibitem [{\citenamefont {Hilfer}\ and\ \citenamefont
  {Wilding}(1995)}]{Hilfer1995}%
  \BibitemOpen
  \bibfield  {author} {\bibinfo {author} {\bibfnamefont {R.}~\bibnamefont
  {Hilfer}}\ and\ \bibinfo {author} {\bibfnamefont {N.}~\bibnamefont
  {Wilding}},\ }\bibfield  {title} {\bibinfo {title} {Are critical finite-size
  scaling functions calculable from knowledge of an appropriate critical
  exponent?},\ }\href {https://doi.org/10.1088/0305-4470/28/10/001} {\bibfield
  {journal} {\bibinfo  {journal} {J. Phys. A: Math. Gen.}\ }\textbf {\bibinfo
  {volume} {28}},\ \bibinfo {pages} {L281} (\bibinfo {year}
  {1995})}\BibitemShut {NoStop}%
\bibitem [{\citenamefont {Smith}\ and\ \citenamefont
  {Bruce}(1995)}]{smith1995study}%
  \BibitemOpen
  \bibfield  {author} {\bibinfo {author} {\bibfnamefont {G.~R.}\ \bibnamefont
  {Smith}}\ and\ \bibinfo {author} {\bibfnamefont {A.~D.}\ \bibnamefont
  {Bruce}},\ }\bibfield  {title} {\bibinfo {title} {{A study of the
  multi-canonical Monte Carlo method}},\ }\href
  {https://doi.org/10.1088/0305-4470/28/23/015} {\bibfield  {journal} {\bibinfo
   {journal} {J. Phys. A: Math. Gen.}\ }\textbf {\bibinfo {volume} {28}},\
  \bibinfo {pages} {6623} (\bibinfo {year} {1995})}\BibitemShut {NoStop}%
\bibitem [{\citenamefont {Tsypin}\ and\ \citenamefont
  {Bl{\"o}te}(2000)}]{tsypin2000}%
  \BibitemOpen
  \bibfield  {author} {\bibinfo {author} {\bibfnamefont {M.~M.}\ \bibnamefont
  {Tsypin}}\ and\ \bibinfo {author} {\bibfnamefont {H.~W.~J.}\ \bibnamefont
  {Bl{\"o}te}},\ }\bibfield  {title} {\bibinfo {title} {{Probability
  distribution of the order parameter for the three-dimensional Ising-model
  universality class: A high-precision Monte Carlo study}},\ }\href
  {https://doi.org/10.1103/PhysRevE.62.73} {\bibfield  {journal} {\bibinfo
  {journal} {Phys. Rev. E}\ }\textbf {\bibinfo {volume} {62}},\ \bibinfo
  {pages} {73} (\bibinfo {year} {2000})}\BibitemShut {NoStop}%
\bibitem [{\citenamefont {Malakis}\ and\ \citenamefont
  {Fytas}(2006)}]{malakis2006universal}%
  \BibitemOpen
  \bibfield  {author} {\bibinfo {author} {\bibfnamefont {A.}~\bibnamefont
  {Malakis}}\ and\ \bibinfo {author} {\bibfnamefont {N.~G.}\ \bibnamefont
  {Fytas}},\ }\bibfield  {title} {\bibinfo {title} {{Universal features and
  tail analysis of the order-parameter distribution of the two-dimensional
  Ising model: An entropic sampling Monte Carlo study}},\ }\href
  {https://doi.org/10.1103/PhysRevE.73.056114} {\bibfield  {journal} {\bibinfo
  {journal} {Phys. Rev. E}\ }\textbf {\bibinfo {volume} {73}},\ \bibinfo
  {pages} {056114} (\bibinfo {year} {2006})}\BibitemShut {NoStop}%
\bibitem [{\citenamefont {Hilfer}\ \emph {et~al.}(2003)\citenamefont {Hilfer},
  \citenamefont {Biswal}, \citenamefont {Mattutis},\ and\ \citenamefont
  {Janke}}]{hilfer2003multicanonical}%
  \BibitemOpen
  \bibfield  {author} {\bibinfo {author} {\bibfnamefont {R.}~\bibnamefont
  {Hilfer}}, \bibinfo {author} {\bibfnamefont {B.}~\bibnamefont {Biswal}},
  \bibinfo {author} {\bibfnamefont {H.~G.}\ \bibnamefont {Mattutis}},\ and\
  \bibinfo {author} {\bibfnamefont {W.}~\bibnamefont {Janke}},\ }\bibfield
  {title} {\bibinfo {title} {{Multicanonical Monte Carlo study and analysis of
  tails for the order-parameter distribution of the two-dimensional Ising
  model}},\ }\href {https://doi.org/10.1103/PhysRevE.68.046123} {\bibfield
  {journal} {\bibinfo  {journal} {Phys. Rev. E}\ }\textbf {\bibinfo {volume}
  {68}},\ \bibinfo {pages} {046123} (\bibinfo {year} {2003})}\BibitemShut
  {NoStop}%
\bibitem [{\citenamefont {Hilfer}\ \emph {et~al.}(2005)\citenamefont {Hilfer},
  \citenamefont {Biswal}, \citenamefont {Mattutis},\ and\ \citenamefont
  {Janke}}]{hilfer2005multicanonical}%
  \BibitemOpen
  \bibfield  {author} {\bibinfo {author} {\bibfnamefont {R.}~\bibnamefont
  {Hilfer}}, \bibinfo {author} {\bibfnamefont {B.}~\bibnamefont {Biswal}},
  \bibinfo {author} {\bibfnamefont {H.-G.}\ \bibnamefont {Mattutis}},\ and\
  \bibinfo {author} {\bibfnamefont {W.}~\bibnamefont {Janke}},\ }\bibfield
  {title} {\bibinfo {title} {{Multicanonical simulations of the tails of the
  order-parameter distribution of the two-dimensional Ising model}},\ }\href
  {https://doi.org/10.1016/j.cpc.2005.03.053} {\bibfield  {journal} {\bibinfo
  {journal} {Comp. Phys. Comm.}\ }\textbf {\bibinfo {volume} {169}},\ \bibinfo
  {pages} {230} (\bibinfo {year} {2005})}\BibitemShut {NoStop}%
\bibitem [{\citenamefont {Balog}\ \emph {et~al.}(2022)\citenamefont {Balog},
  \citenamefont {Ran{\ifmmode \mbox{\c{c}}\else \c{c}\fi{}}on},\ and\
  \citenamefont {Delamotte}}]{Delamotte}%
  \BibitemOpen
  \bibfield  {author} {\bibinfo {author} {\bibfnamefont {I.}~\bibnamefont
  {Balog}}, \bibinfo {author} {\bibfnamefont {A.}~\bibnamefont {Ran{\ifmmode
  \mbox{\c{c}}\else \c{c}\fi{}}on}},\ and\ \bibinfo {author} {\bibfnamefont
  {B.}~\bibnamefont {Delamotte}},\ }\bibfield  {title} {\bibinfo {title}
  {Critical probability distributions of the order parameter from the
  functional renormalization group},\ }\href
  {https://doi.org/10.1103/PhysRevLett.129.210602} {\bibfield  {journal}
  {\bibinfo  {journal} {Phys. Rev. Lett.}\ }\textbf {\bibinfo {volume} {129}},\
  \bibinfo {pages} {210602} (\bibinfo {year} {2022})}\BibitemShut {NoStop}%
\bibitem [{\citenamefont {Balog}\ \emph {et~al.}(2025)\citenamefont {Balog},
  \citenamefont {Delamotte},\ and\ \citenamefont
  {Rancon}}]{balog2025universal}%
  \BibitemOpen
  \bibfield  {author} {\bibinfo {author} {\bibfnamefont {I.}~\bibnamefont
  {Balog}}, \bibinfo {author} {\bibfnamefont {B.}~\bibnamefont {Delamotte}},\
  and\ \bibinfo {author} {\bibfnamefont {A.}~\bibnamefont {Rancon}},\
  }\bibfield  {title} {\bibinfo {title} {Universal and non-universal large
  deviations in critical systems},\ }\href
  {https://doi.org/10.21468/SciPostPhys.18.4.119} {\bibfield  {journal}
  {\bibinfo  {journal} {SciPost Phys.}\ }\textbf {\bibinfo {volume} {18}},\
  \bibinfo {pages} {119} (\bibinfo {year} {2025})}\BibitemShut {NoStop}%
\bibitem [{\citenamefont {Amit}\ and\ \citenamefont
  {Mart\'{i}n-Mayor}(2005)}]{amit:05}%
  \BibitemOpen
  \bibfield  {author} {\bibinfo {author} {\bibfnamefont {D.~J.}\ \bibnamefont
  {Amit}}\ and\ \bibinfo {author} {\bibfnamefont {V.}~\bibnamefont
  {Mart\'{i}n-Mayor}},\ }\href {https://doi.org/10.1142/9789812775313_bmatter}
  {\emph {\bibinfo {title} {Field Theory, the Renormalization Group and
  Critical Phenomena}}},\ \bibinfo {edition} {3rd}\ ed.\ (\bibinfo  {publisher}
  {World Scientific},\ \bibinfo {address} {Singapore},\ \bibinfo {year}
  {2005})\BibitemShut {NoStop}%
\bibitem [{\citenamefont {Cardy}(1984)}]{cardy:84b}%
  \BibitemOpen
  \bibfield  {author} {\bibinfo {author} {\bibfnamefont {J.~L.}\ \bibnamefont
  {Cardy}},\ }\bibfield  {title} {\bibinfo {title} {Conformal invariance and
  universality in finite-size scaling},\ }\href
  {https://doi.org/10.1088/0305-4470/17/7/003} {\bibfield  {journal} {\bibinfo
  {journal} {J. Phys. A: Math. Gen.}\ }\textbf {\bibinfo {volume} {17}},\
  \bibinfo {pages} {L385} (\bibinfo {year} {1984})}\BibitemShut {NoStop}%
\bibitem [{\citenamefont {Burkhardt}\ and\ \citenamefont
  {Derrida}(1985)}]{burkhardt:85}%
  \BibitemOpen
  \bibfield  {author} {\bibinfo {author} {\bibfnamefont {T.~W.}\ \bibnamefont
  {Burkhardt}}\ and\ \bibinfo {author} {\bibfnamefont {B.}~\bibnamefont
  {Derrida}},\ }\bibfield  {title} {\bibinfo {title} {Moments of the total
  magnetization and conformal invariance in the finite two-dimensional ising
  model},\ }\href {https://doi.org/10.1103/PhysRevB.32.7273} {\bibfield
  {journal} {\bibinfo  {journal} {Phys. Rev. B}\ }\textbf {\bibinfo {volume}
  {32}},\ \bibinfo {pages} {7273} (\bibinfo {year} {1985})}\BibitemShut
  {NoStop}%
\bibitem [{\citenamefont {Kaneda}\ and\ \citenamefont
  {Okabe}(2001)}]{kaneda2001finite}%
  \BibitemOpen
  \bibfield  {author} {\bibinfo {author} {\bibfnamefont {K.}~\bibnamefont
  {Kaneda}}\ and\ \bibinfo {author} {\bibfnamefont {Y.}~\bibnamefont {Okabe}},\
  }\bibfield  {title} {\bibinfo {title} {Finite-size scaling for the ising
  model on the m\"obius strip and the klein bottle},\ }\href
  {https://doi.org/10.1103/PhysRevLett.86.2134} {\bibfield  {journal} {\bibinfo
   {journal} {Phys. Rev. Lett.}\ }\textbf {\bibinfo {volume} {86}},\ \bibinfo
  {pages} {2134} (\bibinfo {year} {2001})}\BibitemShut {NoStop}%
\bibitem [{\citenamefont {Bruce}\ and\ \citenamefont
  {Wilding}(1992)}]{bruce1992}%
  \BibitemOpen
  \bibfield  {author} {\bibinfo {author} {\bibfnamefont {A.~D.}\ \bibnamefont
  {Bruce}}\ and\ \bibinfo {author} {\bibfnamefont {N.~B.}\ \bibnamefont
  {Wilding}},\ }\bibfield  {title} {\bibinfo {title} {Scaling fields and
  universality of the liquid-gas critical point},\ }\href
  {https://doi.org/10.1103/PhysRevLett.68.193} {\bibfield  {journal} {\bibinfo
  {journal} {Phys. Rev. Lett.}\ }\textbf {\bibinfo {volume} {68}},\ \bibinfo
  {pages} {193} (\bibinfo {year} {1992})}\BibitemShut {NoStop}%
\bibitem [{\citenamefont {Wilding}(1997)}]{wilding:1997}%
  \BibitemOpen
  \bibfield  {author} {\bibinfo {author} {\bibfnamefont {N.~B.}\ \bibnamefont
  {Wilding}},\ }\bibfield  {title} {\bibinfo {title} {Simulation studies of
  fluid critical behaviour},\ }\href
  {https://doi.org/10.1088/0953-8984/9/3/002} {\bibfield  {journal} {\bibinfo
  {journal} {J. Phys. Condens. Matter}\ }\textbf {\bibinfo {volume} {9}},\
  \bibinfo {pages} {585} (\bibinfo {year} {1997})}\BibitemShut {NoStop}%
\bibitem [{\citenamefont {Touchette}(2009)}]{Touchette2009}%
  \BibitemOpen
  \bibfield  {author} {\bibinfo {author} {\bibfnamefont {H.}~\bibnamefont
  {Touchette}},\ }\bibfield  {title} {\bibinfo {title} {The large deviation
  approach to statistical mechanics},\ }\href
  {https://doi.org/https://doi.org/10.1016/j.physrep.2009.05.002} {\bibfield
  {journal} {\bibinfo  {journal} {Phys. Rep.}\ }\textbf {\bibinfo {volume}
  {478}},\ \bibinfo {pages} {1} (\bibinfo {year} {2009})}\BibitemShut {NoStop}%
\bibitem [{\citenamefont {Stella}\ \emph {et~al.}(2023)\citenamefont {Stella},
  \citenamefont {Chechkin},\ and\ \citenamefont {Teza}}]{stella2023anomalous}%
  \BibitemOpen
  \bibfield  {author} {\bibinfo {author} {\bibfnamefont {A.~L.}\ \bibnamefont
  {Stella}}, \bibinfo {author} {\bibfnamefont {A.}~\bibnamefont {Chechkin}},\
  and\ \bibinfo {author} {\bibfnamefont {G.}~\bibnamefont {Teza}},\ }\bibfield
  {title} {\bibinfo {title} {{Anomalous Dynamical Scaling Determines Universal
  Critical Singularities}},\ }\href
  {https://doi.org/10.1103/PhysRevLett.130.207104} {\bibfield  {journal}
  {\bibinfo  {journal} {Phys. Rev. Lett.}\ }\textbf {\bibinfo {volume} {130}},\
  \bibinfo {pages} {207104} (\bibinfo {year} {2023})}\BibitemShut {NoStop}%
\bibitem [{\citenamefont {McCoy}\ and\ \citenamefont
  {Wu}(1973)}]{mccoy1973two}%
  \BibitemOpen
  \bibfield  {author} {\bibinfo {author} {\bibfnamefont {B.~M.}\ \bibnamefont
  {McCoy}}\ and\ \bibinfo {author} {\bibfnamefont {T.~T.}\ \bibnamefont {Wu}},\
  }\href@noop {} {\emph {\bibinfo {title} {The Two-Dimensional {I}sing
  Model}}}\ (\bibinfo  {publisher} {Harvard University Press},\ \bibinfo
  {address} {Cambridge, Massachusetts},\ \bibinfo {year} {1973})\BibitemShut
  {NoStop}%
\bibitem [{\citenamefont {{F. Camia, C. Garban, and C.M.
  Newman}}(2016)}]{camia2016planar}%
  \BibitemOpen
  \bibfield  {author} {\bibinfo {author} {\bibnamefont {{F. Camia, C. Garban,
  and C.M. Newman}}},\ }\bibfield  {title} {\bibinfo {title} {{Planar Ising
  magnetization field II. Properties of the critical and near-critical scaling
  limits}},\ }\href {https://doi.org/10.1214/14-AIHP643} {\bibfield  {journal}
  {\bibinfo  {journal} {Ann. Inst. Henri Poincaré Probab. Stat.}\ }\textbf
  {\bibinfo {volume} {52}},\ \bibinfo {pages} {146 } (\bibinfo {year}
  {2016})}\BibitemShut {NoStop}%
\bibitem [{\citenamefont {Camia}\ \emph {et~al.}(2025)\citenamefont {Camia},
  \citenamefont {Dakkak},\ and\ \citenamefont {Peccati}}]{camia2025power}%
  \BibitemOpen
  \bibfield  {author} {\bibinfo {author} {\bibfnamefont {F.}~\bibnamefont
  {Camia}}, \bibinfo {author} {\bibfnamefont {O.~E.}\ \bibnamefont {Dakkak}},\
  and\ \bibinfo {author} {\bibfnamefont {G.}~\bibnamefont {Peccati}},\
  }\bibfield  {title} {\bibinfo {title} {{Power-law correction in the
  probability density function of the critical Ising magnetization}},\ }\href
  {https://doi.org/10.1007/s11040-025-09517-z} {\bibfield  {journal} {\bibinfo
  {journal} {Math. Phys. Anal. Geom.}\ }\textbf {\bibinfo {volume} {28}},\
  \bibinfo {pages} {1} (\bibinfo {year} {2025})}\BibitemShut {NoStop}%
\bibitem [{\citenamefont {Blume}(1966)}]{blume}%
  \BibitemOpen
  \bibfield  {author} {\bibinfo {author} {\bibfnamefont {M.}~\bibnamefont
  {Blume}},\ }\bibfield  {title} {\bibinfo {title} {{Theory of the First-Order
  Magnetic Phase Change in U${\mathrm{O}}_{2}$}},\ }\href
  {https://doi.org/10.1103/PhysRev.141.517} {\bibfield  {journal} {\bibinfo
  {journal} {Phys. Rev.}\ }\textbf {\bibinfo {volume} {141}},\ \bibinfo {pages}
  {517} (\bibinfo {year} {1966})}\BibitemShut {NoStop}%
\bibitem [{\citenamefont {Capel}(1966)}]{capel6}%
  \BibitemOpen
  \bibfield  {author} {\bibinfo {author} {\bibfnamefont {H.~W.}\ \bibnamefont
  {Capel}},\ }\bibfield  {title} {\bibinfo {title} {{On the possibility of
  first-order phase transitions in Ising systems of triplet ions with
  zero-field splitting}},\ }\href
  {https://doi.org/https://doi.org/10.1016/0031-8914(66)90027-9} {\bibfield
  {journal} {\bibinfo  {journal} {Physica}\ }\textbf {\bibinfo {volume} {32}},\
  \bibinfo {pages} {966} (\bibinfo {year} {1966})}\BibitemShut {NoStop}%
\bibitem [{\citenamefont {Fernández}\ \emph {et~al.}(2009)\citenamefont
  {Fernández}, \citenamefont {Martín-Mayor},\ and\ \citenamefont
  {Yllanes}}]{fernandez:09}%
  \BibitemOpen
  \bibfield  {author} {\bibinfo {author} {\bibfnamefont {L.}~\bibnamefont
  {Fernández}}, \bibinfo {author} {\bibfnamefont {V.}~\bibnamefont
  {Martín-Mayor}},\ and\ \bibinfo {author} {\bibfnamefont {D.}~\bibnamefont
  {Yllanes}},\ }\bibfield  {title} {\bibinfo {title} {{Tethered Monte Carlo:
  Computing the effective potential without critical slowing down}},\ }\href
  {https://doi.org/https://doi.org/10.1016/j.nuclphysb.2008.07.009} {\bibfield
  {journal} {\bibinfo  {journal} {Nucl. Phys. B}\ }\textbf {\bibinfo {volume}
  {807}},\ \bibinfo {pages} {424} (\bibinfo {year} {2009})}\BibitemShut
  {NoStop}%
\bibitem [{\citenamefont {Mart\'{i}n-Mayor}\ \emph {et~al.}(2011)\citenamefont
  {Mart\'{i}n-Mayor}, \citenamefont {Seoane},\ and\ \citenamefont
  {Yllanes}}]{martin-mayor:11}%
  \BibitemOpen
  \bibfield  {author} {\bibinfo {author} {\bibfnamefont {V.}~\bibnamefont
  {Mart\'{i}n-Mayor}}, \bibinfo {author} {\bibfnamefont {B.}~\bibnamefont
  {Seoane}},\ and\ \bibinfo {author} {\bibfnamefont {D.}~\bibnamefont
  {Yllanes}},\ }\bibfield  {title} {\bibinfo {title} {{Tethered Monte Carlo:
  Managing Rugged Free-Energy Landscapes with a Helmholtz-Potential
  Formalism}},\ }\href {https://doi.org/10.1007/s10955-011-0261-4} {\bibfield
  {journal} {\bibinfo  {journal} {J. Stat. Phys.}\ }\textbf {\bibinfo {volume}
  {144}},\ \bibinfo {pages} {554} (\bibinfo {year} {2011})}\BibitemShut
  {NoStop}%
\bibitem [{\citenamefont {Zhu}\ \emph {et~al.}(2026)\citenamefont {Zhu},
  \citenamefont {Xu}, \citenamefont {Saberi},\ and\ \citenamefont
  {Deng}}]{zhu2026}%
  \BibitemOpen
  \bibfield  {author} {\bibinfo {author} {\bibfnamefont {J.}~\bibnamefont
  {Zhu}}, \bibinfo {author} {\bibfnamefont {Y.}~\bibnamefont {Xu}}, \bibinfo
  {author} {\bibfnamefont {A.~A.}\ \bibnamefont {Saberi}},\ and\ \bibinfo
  {author} {\bibfnamefont {Y.}~\bibnamefont {Deng}},\ }\href
  {https://arxiv.org/abs/2606.19030} {\bibinfo {title} {{Large-deviation tails
  of critical order-parameter distributions}}} (\bibinfo {year} {2026}),\
  \Eprint {https://arxiv.org/abs/2606.19030} {arXiv:2606.19030
  [cond-mat.stat-mech]} \BibitemShut {NoStop}%
\bibitem [{\citenamefont {Salas}\ and\ \citenamefont {Sokal}(2000)}]{salas:00}%
  \BibitemOpen
  \bibfield  {author} {\bibinfo {author} {\bibfnamefont {J.}~\bibnamefont
  {Salas}}\ and\ \bibinfo {author} {\bibfnamefont {A.~D.}\ \bibnamefont
  {Sokal}},\ }\bibfield  {title} {\bibinfo {title} {{Universal Amplitude Ratios
  in the Critical Two-Dimensional Ising Model on a Torus}},\ }\href
  {https://doi.org/10.1023/A:1018611122166} {\bibfield  {journal} {\bibinfo
  {journal} {J. Stat. Phys.}\ }\textbf {\bibinfo {volume} {98}},\ \bibinfo
  {pages} {551} (\bibinfo {year} {2000})}\BibitemShut {NoStop}%
\bibitem [{\citenamefont {Mart\'{i}n-Mayor}(2007)}]{martin-mayor:07}%
  \BibitemOpen
  \bibfield  {author} {\bibinfo {author} {\bibfnamefont {V.}~\bibnamefont
  {Mart\'{i}n-Mayor}},\ }\bibfield  {title} {\bibinfo {title} {{Microcanonical
  Approach to the Simulation of First-Order Phase Transitions}},\ }\href
  {https://doi.org/10.1103/PhysRevLett.98.137207} {\bibfield  {journal}
  {\bibinfo  {journal} {Phys. Rev. Lett.}\ }\textbf {\bibinfo {volume} {98}},\
  \bibinfo {pages} {137207} (\bibinfo {year} {2007})}\BibitemShut {NoStop}%
\bibitem [{\citenamefont {Mart\'{i}n-Mayor}\ and\ \citenamefont
  {Yllanes}(2009)}]{martin-mayor:09}%
  \BibitemOpen
  \bibfield  {author} {\bibinfo {author} {\bibfnamefont {V.}~\bibnamefont
  {Mart\'{i}n-Mayor}}\ and\ \bibinfo {author} {\bibfnamefont {D.}~\bibnamefont
  {Yllanes}},\ }\bibfield  {title} {\bibinfo {title} {{A cluster {M}onte
  {C}arlo algorithm with a conserved order parameter}},\ }\href
  {https://doi.org/10.1103/PhysRevE.80.015701} {\bibfield  {journal} {\bibinfo
  {journal} {Phys. Rev. E}\ }\textbf {\bibinfo {volume} {80}},\ \bibinfo
  {pages} {015701(R)} (\bibinfo {year} {2009})}\BibitemShut {NoStop}%
\bibitem [{\citenamefont {Yllanes}(2011)}]{yllanes:11}%
  \BibitemOpen
  \bibfield  {author} {\bibinfo {author} {\bibfnamefont {D.}~\bibnamefont
  {Yllanes}},\ }\emph {\bibinfo {title} {{Rugged Free-Energy Landscapes in
  Disordered Spin Systems}}},\ \href@noop {} {Ph.D. thesis},\ \bibinfo
  {school} {Universidad Complutense de Madrid} (\bibinfo {year} {2011}),\
  \Eprint {https://arxiv.org/abs/arXiv:1111.0266} {arXiv:1111.0266}
  \BibitemShut {NoStop}%
\bibitem [{\citenamefont {Caselle}\ and\ \citenamefont
  {Hasenbusch}(2000)}]{caselle:00b}%
  \BibitemOpen
  \bibfield  {author} {\bibinfo {author} {\bibfnamefont {M.}~\bibnamefont
  {Caselle}}\ and\ \bibinfo {author} {\bibfnamefont {M.}~\bibnamefont
  {Hasenbusch}},\ }\bibfield  {title} {\bibinfo {title} {{Critical amplitudes
  and mass spectrum of the 2d Ising model in a magnetic field}},\ }\href
  {https://doi.org/https://doi.org/10.1016/S0550-3213(00)00074-2} {\bibfield
  {journal} {\bibinfo  {journal} {Nucl. Phys. B}\ }\textbf {\bibinfo {volume}
  {579}},\ \bibinfo {pages} {667} (\bibinfo {year} {2000})}\BibitemShut
  {NoStop}%
\bibitem [{\citenamefont {Bisson}\ \emph {et~al.}(2025)\citenamefont {Bisson},
  \citenamefont {Vasilopoulos}, \citenamefont {Bernaschi}, \citenamefont
  {Fatica}, \citenamefont {Fytas}, \citenamefont {Pemart\'{\i}n},\ and\
  \citenamefont {Mart\'{\i}n-Mayor}}]{bisson:25}%
  \BibitemOpen
  \bibfield  {author} {\bibinfo {author} {\bibfnamefont {M.}~\bibnamefont
  {Bisson}}, \bibinfo {author} {\bibfnamefont {A.}~\bibnamefont
  {Vasilopoulos}}, \bibinfo {author} {\bibfnamefont {M.}~\bibnamefont
  {Bernaschi}}, \bibinfo {author} {\bibfnamefont {M.}~\bibnamefont {Fatica}},
  \bibinfo {author} {\bibfnamefont {N.~G.}\ \bibnamefont {Fytas}}, \bibinfo
  {author} {\bibfnamefont {I.~G.-A.}\ \bibnamefont {Pemart\'{\i}n}},\ and\
  \bibinfo {author} {\bibfnamefont {V.}~\bibnamefont {Mart\'{\i}n-Mayor}},\
  }\bibfield  {title} {\bibinfo {title} {{Universal exotic dynamics in critical
  mesoscopic systems: Simulating the square root of Avogadro's number of
  spins}},\ }\href {https://doi.org/10.1103/ngkf-7816} {\bibfield  {journal}
  {\bibinfo  {journal} {Phys. Rev. Res.}\ }\textbf {\bibinfo {volume} {7}},\
  \bibinfo {pages} {033218} (\bibinfo {year} {2025})}\BibitemShut {NoStop}%
\bibitem [{\citenamefont {Caselle}\ \emph {et~al.}(2002)\citenamefont
  {Caselle}, \citenamefont {Hasenbusch}, \citenamefont {Pelissetto},\ and\
  \citenamefont {Vicari}}]{caselle:02}%
  \BibitemOpen
  \bibfield  {author} {\bibinfo {author} {\bibfnamefont {M.}~\bibnamefont
  {Caselle}}, \bibinfo {author} {\bibfnamefont {M.}~\bibnamefont {Hasenbusch}},
  \bibinfo {author} {\bibfnamefont {A.}~\bibnamefont {Pelissetto}},\ and\
  \bibinfo {author} {\bibfnamefont {E.}~\bibnamefont {Vicari}},\ }\bibfield
  {title} {\bibinfo {title} {{Irrelevant operators in the two-dimensional Ising
  model}},\ }\href {https://doi.org/10.1088/0305-4470/35/23/305} {\bibfield
  {journal} {\bibinfo  {journal} {J. Phys. A: Math. Gen.}\ }\textbf {\bibinfo
  {volume} {35}},\ \bibinfo {pages} {4861} (\bibinfo {year}
  {2002})}\BibitemShut {NoStop}%
\bibitem [{\citenamefont {Jeon}\ \emph {et~al.}(2026)\citenamefont {Jeon},
  \citenamefont {Vasilopoulos}, \citenamefont {Kim}, \citenamefont
  {Martín-Mayor},\ and\ \citenamefont {Fytas}}]{jeon2026}%
  \BibitemOpen
  \bibfield  {author} {\bibinfo {author} {\bibfnamefont {M.}~\bibnamefont
  {Jeon}}, \bibinfo {author} {\bibfnamefont {A.}~\bibnamefont {Vasilopoulos}},
  \bibinfo {author} {\bibfnamefont {D.-H.}\ \bibnamefont {Kim}}, \bibinfo
  {author} {\bibfnamefont {V.}~\bibnamefont {Martín-Mayor}},\ and\ \bibinfo
  {author} {\bibfnamefont {N.~G.}\ \bibnamefont {Fytas}},\ }\href
  {https://arxiv.org/abs/2605.07867} {\bibinfo {title} {{Cluster dynamics stay
  fast until tricriticality}}} (\bibinfo {year} {2026}),\ \Eprint
  {https://arxiv.org/abs/2605.07867} {arXiv:2605.07867 [cond-mat.stat-mech]}
  \BibitemShut {NoStop}%
\bibitem [{\citenamefont {Parisi}(1988)}]{parisi:88}%
  \BibitemOpen
  \bibfield  {author} {\bibinfo {author} {\bibfnamefont {G.}~\bibnamefont
  {Parisi}},\ }\href@noop {} {\emph {\bibinfo {title} {Statistical Field
  Theory}}}\ (\bibinfo  {publisher} {Addison-Wesley},\ \bibinfo {address}
  {Redwood City, CA},\ \bibinfo {year} {1988})\BibitemShut {NoStop}%
\bibitem [{\citenamefont {Bouchaud}\ and\ \citenamefont
  {Georges}(1990)}]{bouchaud:90}%
  \BibitemOpen
  \bibfield  {author} {\bibinfo {author} {\bibfnamefont {J.-P.}\ \bibnamefont
  {Bouchaud}}\ and\ \bibinfo {author} {\bibfnamefont {A.}~\bibnamefont
  {Georges}},\ }\bibfield  {title} {\bibinfo {title} {Anomalous diffusion in
  disordered media: Statistical mechanisms, models and physical applications},\
  }\href {https://doi.org/https://doi.org/10.1016/0370-1573(90)90099-N}
  {\bibfield  {journal} {\bibinfo  {journal} {Phys. Rep.}\ }\textbf {\bibinfo
  {volume} {195}},\ \bibinfo {pages} {127} (\bibinfo {year}
  {1990})}\BibitemShut {NoStop}%
\bibitem [{\citenamefont {Barber}(1983)}]{barber:83}%
  \BibitemOpen
  \bibfield  {author} {\bibinfo {author} {\bibfnamefont {M.~N.}\ \bibnamefont
  {Barber}},\ }\bibfield  {title} {\bibinfo {title} {Finite-size scaling}\
  }(\bibinfo  {publisher} {Academic Press},\ \bibinfo {year}
  {1983})\BibitemShut {NoStop}%
\bibitem [{\citenamefont {Chihara}(1978)}]{chihara:78}%
  \BibitemOpen
  \bibfield  {author} {\bibinfo {author} {\bibfnamefont {T.}~\bibnamefont
  {Chihara}},\ }\href@noop {} {\emph {\bibinfo {title} {An Introduction to
  Orthogonal Polynomials}}}\ (\bibinfo  {publisher} {Gordon \& Breach},\
  \bibinfo {address} {New York},\ \bibinfo {year} {1978})\BibitemShut {NoStop}%
\bibitem [{\citenamefont {{Di Francesco}}\ \emph {et~al.}(1987)\citenamefont
  {{Di Francesco}}, \citenamefont {Saleur},\ and\ \citenamefont
  {Zuber}}]{difrancesco:87}%
  \BibitemOpen
  \bibfield  {author} {\bibinfo {author} {\bibfnamefont {P.}~\bibnamefont {{Di
  Francesco}}}, \bibinfo {author} {\bibfnamefont {H.}~\bibnamefont {Saleur}},\
  and\ \bibinfo {author} {\bibfnamefont {J.}~\bibnamefont {Zuber}},\ }\bibfield
   {title} {\bibinfo {title} {{Critical Ising correlation functions in the
  plane and on the torus}},\ }\href
  {https://doi.org/https://doi.org/10.1016/0550-3213(87)90202-1} {\bibfield
  {journal} {\bibinfo  {journal} {Nucl. Phys. B}\ }\textbf {\bibinfo {volume}
  {290}},\ \bibinfo {pages} {527} (\bibinfo {year} {1987})}\BibitemShut
  {NoStop}%
\bibitem [{\citenamefont {{Di Francesco}}\ \emph {et~al.}(1988)\citenamefont
  {{Di Francesco}}, \citenamefont {Saleur},\ and\ \citenamefont
  {Zuber}}]{difrancesco:88}%
  \BibitemOpen
  \bibfield  {author} {\bibinfo {author} {\bibfnamefont {P.}~\bibnamefont {{Di
  Francesco}}}, \bibinfo {author} {\bibfnamefont {H.}~\bibnamefont {Saleur}},\
  and\ \bibinfo {author} {\bibfnamefont {J.}~\bibnamefont {Zuber}},\ }\bibfield
   {title} {\bibinfo {title} {{Correlation Functions of the Critical Ising
  Model on a Torus}},\ }\href {https://doi.org/10.1209/0295-5075/5/2/001}
  {\bibfield  {journal} {\bibinfo  {journal} {Europhys. Lett.}\ }\textbf
  {\bibinfo {volume} {5}},\ \bibinfo {pages} {95} (\bibinfo {year}
  {1988})}\BibitemShut {NoStop}%
\bibitem [{\citenamefont {Ferdinand}\ and\ \citenamefont
  {Fisher}(1969)}]{ferdinand:69}%
  \BibitemOpen
  \bibfield  {author} {\bibinfo {author} {\bibfnamefont {A.~E.}\ \bibnamefont
  {Ferdinand}}\ and\ \bibinfo {author} {\bibfnamefont {M.~E.}\ \bibnamefont
  {Fisher}},\ }\bibfield  {title} {\bibinfo {title} {{Bounded and Inhomogeneous
  Ising Models. I. Specific-Heat Anomaly of a Finite Lattice}},\ }\href
  {https://doi.org/10.1103/PhysRev.185.832} {\bibfield  {journal} {\bibinfo
  {journal} {Phys. Rev.}\ }\textbf {\bibinfo {volume} {185}},\ \bibinfo {pages}
  {832} (\bibinfo {year} {1969})}\BibitemShut {NoStop}%
\bibitem [{\citenamefont {Sokal}(1997)}]{sokal:97}%
  \BibitemOpen
  \bibfield  {author} {\bibinfo {author} {\bibfnamefont {A.~D.}\ \bibnamefont
  {Sokal}},\ }\bibfield  {title} {\bibinfo {title} {{Monte Carlo Methods in
  Statistical Mechanics: Foundations and New Algorithms}},\ }in\ \href
  {https://doi.org/10.1007/978-1-4899-0319-8_6} {\emph {\bibinfo {booktitle}
  {Functional Integration: Basics and Applications (1996 Carg{\`e}se
  School)}}},\ \bibinfo {editor} {edited by\ \bibinfo {editor} {\bibfnamefont
  {C.}~\bibnamefont {DeWitt-Morette}}, \bibinfo {editor} {\bibfnamefont
  {P.}~\bibnamefont {Cartier}},\ and\ \bibinfo {editor} {\bibfnamefont
  {A.}~\bibnamefont {Folacci}}}\ (\bibinfo  {publisher} {Plenum},\ \bibinfo
  {address} {N. Y.},\ \bibinfo {year} {1997})\BibitemShut {NoStop}%
\end{thebibliography}%

\end{document}